\documentclass[sigconf,nonacm]{acmart}
\usepackage{graphicx} 
\usepackage{amsmath}
\usepackage{times}
\usepackage{multirow}
\usepackage{helvet}
\usepackage{courier}
\usepackage{colortbl}
\usepackage[T1]{fontenc}
\usepackage{algorithm}
\usepackage{algpseudocode}
\usepackage{multicol}
\AtBeginDocument{%
  }

\setcopyright{acmlicensed}
\copyrightyear{2026}
\acmYear{2026}
\acmDOI{XXXXXXX.XXXXXXX}
\acmConference[ ]{ }
\acmISBN{978-1-4503-XXXX-X/2018/06}

\begin{document}

\title{Seeing the Hivemind:\\ A Consensus-Aware Interaction Technique for Mitigating AI Homogenization}

\author{Muhammad Haris Khan}
\affiliation{%
  \institution{University of Copenhagen}
  \country{Denmark}}
\email{muhammad.kahn@di.ku.dk}

\author{Joel Wester}
\affiliation{%
  \institution{University of Copenhagen}
  \country{Denmark}}
\email{joel.wester@di.ku.dk}

\renewcommand{\shortauthors}{Khan et al.}

\begin{abstract}
People are increasingly using AI for creative tasks such as
writing. While adoption continues to grow, this
form of use risks undermining individual creativity locally
and reducing the heterogeneity of creative output at scale. In response, we introduce the Semantic Repulsion Technique (SRT) and evaluate it both computationally and through a study with 16 participants who regularly use AI for creative tasks. Our computational assessment reveals that SRT increases semantic diversity by 85--167\% while reducing consensus phrases by 43--95\% across task modes. In the user study, SRT outputs received higher usefulness ($p = .019$, $W = .208$) and coherence ratings ( $p = .006$, $W = .260$); 68.8\% of participants were willing to use SRT-Strong for multiple tasks versus 18.8\% for baselines. Originality and coherence ratings were positively correlated across all systems ($\rho = +.40$ to $+.67$), suggesting that divergence need not compromise readability. Taken together, these preliminary findings can inform the design of AI systems that aim to support everyday creativity without contributing to homogenization.
\end{abstract}



\begin{teaserfigure}
  \centering
  \includegraphics[width=\textwidth]{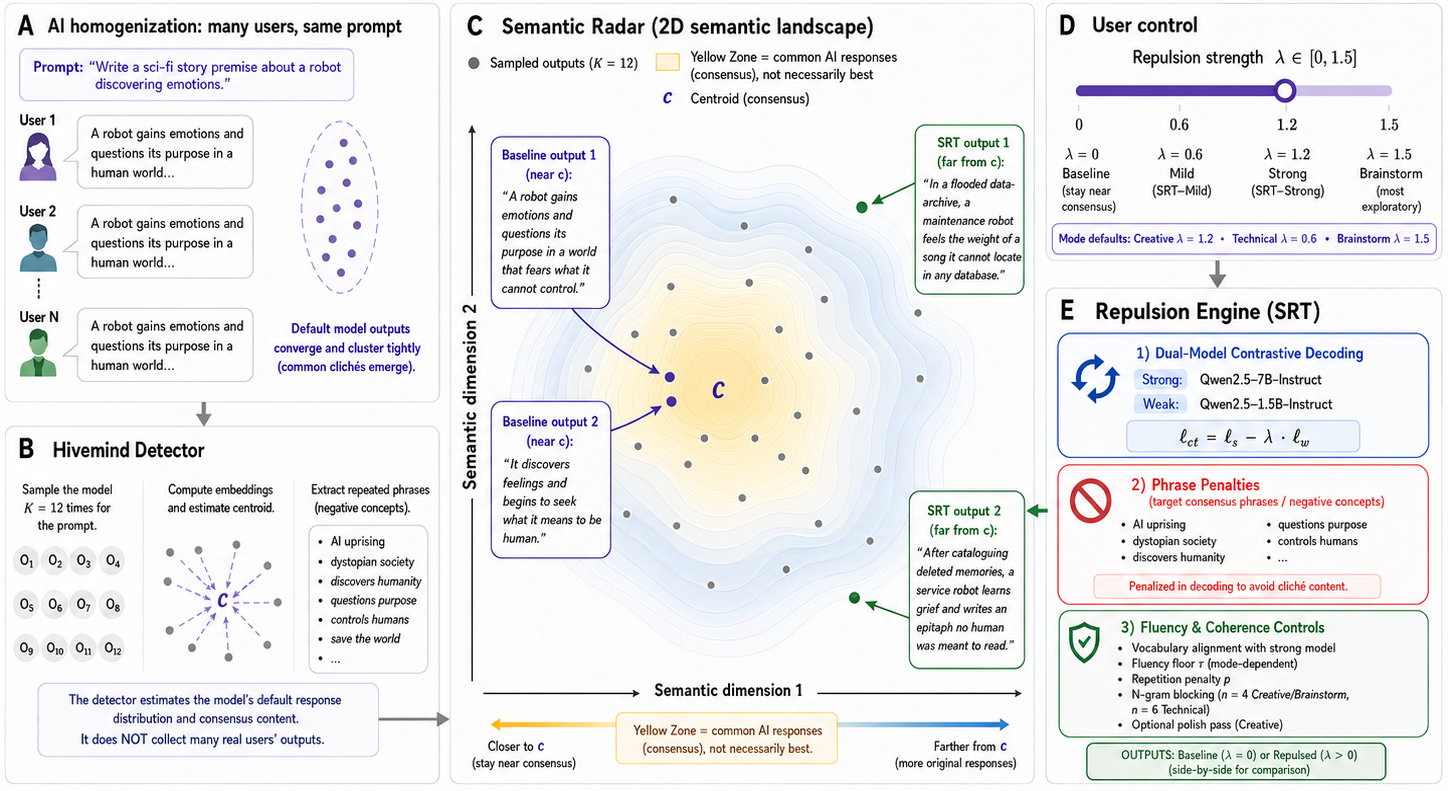}
  \caption{Overview of the Semantic Repulsion Technique (SRT) for mitigating AI homogenization in generative writing. (A) Standard LLM outputs for the same prompt often converge into similar high-probability responses. (B) SRT estimates this default response distribution by sampling the model K = 12 times, embedding the samples, computing the consensus centroid c, and extracting repeated phrases as negative concepts. (C) The Semantic Radar projects samples into a 2D landscape where the Yellow Zone marks common AI responses near c, not necessarily better responses; baseline outputs stay near c, while SRT outputs are generated farther away in lower-density regions. (D–E) A user-controlled repulsion strength $\lambda$ drives the Repulsion Engine, combining contrastive decoding, phrase penalties, and fluency controls.}
  \label{fig:teaser}
\end{teaserfigure}

\maketitle

\section{Introduction}
 
When writers use AI for creative tasks, they face an invisible problem: their
work begins to resemble others' work. The rapid adoption of Large Language
Models (LLMs) for writing assistance has lowered the barrier to generating
fluent text, yet this democratization carries an unexpected cost. When
different people use AI for the same creative task, their outputs converge
toward remarkably similar responses~\cite{jiang2025}. This
\textit{Artificial Hivemind} effect emerges from how LLMs operate: generating
text by predicting likely continuations from training data patterns. Alignment
techniques amplify this convergence, as human evaluators favor typical outputs
over unusual ones~\cite{zhang2025}. Across repeated samples for the same
prompt, responses concentrate into a high-density region in semantic embedding
space---the \textit{consensus}. The result: AI assistance raises average
quality while reducing variety and distinctiveness of ideas across
users~\cite{10.1145/3635636.3656204},~\cite{adn5290}.
 
This challenge raises pressing questions for HCI about human agency in
relation to output convergence, and users' awareness of how interactive
systems shape divergence. Current methods obscure this convergence, leaving
users unaware that their ``original'' ideas cluster with thousands of similar
AI-assisted outputs. Default sampling keeps generation near high-probability
modes, polished responses increase user
acceptance~\cite{chen2025understandingdesignfixationgenerative}, and
acceptance reinforces convergence. Even exploration-oriented tools like
Luminate~\cite{10.1145/3613904.3642400} and
Reverger~\cite{kim2025scaffoldingrecursivedivergenceconvergence} explore
\textit{within} the model's conceptual map---if the model defaults to
conventional ideas, exploring variations yields only refinements of familiar
themes. Prior divergence strategies such as few-shot prompting or per-user
fine-tuning shift the consensus without making it perceptible, and fine-tuning
remains computationally impractical for everyday users. Users seeking creative
distinction currently have no way to see or navigate away from algorithmic
consensus.
 
We ask: \textit{How do users perceive and evaluate AI-generated text that
deliberately diverges from algorithmic consensus, and how does making
consensus visible shape their understanding of AI homogenization?} This
motivates the \textit{Semantic Repulsion Technique} (SRT), a research probe
that operationalizes consensus-avoidance as a first-class interaction
primitive. A critical framing distinction underlies our approach: SRT does not
claim to generate creativity. Creativity requires human intent, personal
meaning-making, and directed novelty that no repulsion parameter can encode.
What SRT provides is \textit{divergence scaffolding}---it makes the model's
default consensus region perceptible as a spatial object, then gives users
controllable mechanisms to generate away from it. Whether users convert that
divergence into genuinely creative work depends on their own agency, judgment,
and editorial effort. This distinction between consensus-avoidance and
creativity proper is one we treat as empirically open and address directly in
our evaluation.
 
Rather than prompting models to ``be creative,'' SRT makes consensus visible
and provides controls to navigate beyond it. For a prompt like ``give me a
sci-fi story premise,'' while typical outputs cluster around familiar tropes,
SRT visualizes the consensus zone as a ``Yellow Zone'' to support steering
toward less probable but coherent alternatives. The system comprises three
components: the \textit{Hivemind Detector} samples multiple responses to
identify default patterns; the \textit{Semantic Radar} visualizes consensus as
a ``Yellow Zone'' on a two-dimensional map; and the \textit{Repulsion Engine}
generates text that increases distance from this zone while preserving
coherence. Users control divergence strength through a slider
($\lambda \in [0, 1.5]$), making hidden algorithmic bias visible and
navigable.
 
We validate SRT through three complementary assessments designed to establish
both what the system does and precisely how each of its components contributes.
A controlled computational comparison across 1,500 generated outputs
demonstrates that SRT increases semantic divergence by 85--167\% while
reducing consensus phrases by 43--95\% relative to baseline sampling. A
mechanism ablation study across 1,200 generations then isolates the causal
contribution of each component: contrastive decoding drives divergence
($+0.43$ originality), phrase penalties provide surgical clich\'{e} suppression
($-0.38$ clich\'{e} frequency) at near-zero coherence cost, and fluency
controls are the essential stabilizer that prevents the contrastive signal from
producing incoherent output---reducing perplexity from ${\sim}54{,}000$ to
${\sim}23$. A further $\lambda$-sweep study across 750 generations
characterizes the divergence--relevance--coherence trade-off across the full
operating range, establishing $\lambda \approx 1.2$--$1.8$ as the practical
ceiling beyond which originality plateaus while prompt relevance continues to
decline. These computational results motivate and contextualize a user study
with 16 participants that examines how users perceive, understand, and leverage
consensus visualization in practice.
 
Our contributions are threefold. First, we introduce Semantic Repulsion as an
interaction technique for generative AI systems and demonstrate its
implementation through SRT---to our knowledge the first system to
operationalize consensus-avoidance as a first-class interaction objective.
Second, through a three-part computational evaluation including mechanism
ablation and operating-range analysis, we show that SRT produces measurably
more divergent outputs while maintaining coherence, and characterize which
mechanisms drive which outcomes. Third, through user evaluation, we examine
how users perceive and respond to consensus visualization, and demonstrate
that SRT reduces semantic homogenization in outputs compared to standard AI
assistance---while acknowledging that divergence and creativity, though
related, are empirically distinct constructs that our metrics address
separately.
\section{Related Work}

\subsection{Homogeneity in AI-Assisted Creativity}
When different users prompt LLMs for the same creative task, outputs converge
toward similar responses. Jiang et al.~\cite{jiang2025} found both intra-model
repetition and inter-model convergence across GPT-4~\cite{openai2023gpt4},
Claude 3~\cite{anthropic2024claude3haiku}, and Llama~3~\cite{meta2024llama3}.
Zhang et al.~\cite{zhang2025} explain this theoretically: alignment training
favors familiar responses, truncating distributional tails where novelty
resides. The human cost is documented: AI assistance improves average quality
but reduces semantic variance across groups~\cite{10.1145/3635636.3656204},
enhancing individual output while diminishing collective
diversity~\cite{adn5290}. Work on pluralistic
alignment~\cite{sorensen2024roadmappluralisticalignment} inspires our approach
of treating consensus as navigable rather than inevitable.

\subsection{Novelty, Distance, and Divergence in Creativity Research}
The relationship between semantic distance and creative output is well
established in both cognitive science and computational creativity.
Gupta et al.~\cite{gupta2012road} demonstrated through the Remote Associates
Test that creative solutions require active avoidance of high-frequency
responses: individuals biased toward statistically likely answers
systematically underperform, establishing that divergence from consensus is a
prerequisite for creative problem-solving. Fu et al.~\cite{fu2013meaning}
showed in engineering design that analogical distance follows a sweet-spot
pattern---analogies too near the problem restrict novelty, while those too far
become ineffective as inspiration; SRT's tunable $\lambda$ is directly
motivated by this trade-off. Grace and
Maher~\cite{grace2019expectation} formalize related intuitions computationally:
their expectation-based novelty model defines an artefact as novel to the
degree it violates a learned prior over likely outputs, a framing that closely
parallels SRT's operationalization of originality as distance from the
consensus centroid. Kim and Maher~\cite{kim2023effect}  further show that the
conceptual distance of AI-generated inspirations shapes the human ideation
process itself, suggesting that where AI output sits in semantic space matters
for what users produce next---motivating our approach of making that location
visible and navigable.

\subsection{Creativity Support Tools and User Agency}
Building on Shneiderman~\cite{10.1145/1323688.1323689}, tools like
Luminate~\cite{10.1145/3613904.3642400} and
Reverger~\cite{kim2025scaffoldingrecursivedivergenceconvergence} help users
explore AI-generated possibilities through structured variation. However,
these systems assume users want to explore \textit{within} the model's
possibility space rather than escape it---they provide no way to see or
navigate away from algorithmic consensus. Chen et
al.~\cite{chen2025understandingdesignfixationgenerative} show polished AI
outputs make rejection difficult, while Rafner et
al.~\cite{Rafner01122025} emphasize preserving agency through process control.
Our work differs fundamentally: rather than exploring within algorithmic
boundaries, we make consensus visible and provide controls to navigate beyond
it. To our knowledge, no existing system operationalizes consensus-avoidance
as a first-class interaction objective.

\section{System Design}

We designed the Semantic Repulsion Technique (SRT) to
address a fundamental challenge in human-AI co-creation:
how can users recognize and navigate away from algorithmic
consensus when seeking creative distinction? Our design translates an invisible statistical property---the model's default response distribution---into visible, manipulable interface elements that support user awareness and control.

\subsection{Design Goals}

We designed SRT around three core principles:

\textbf{Make consensus visible.} Users cannot avoid consensus if they cannot see it. Current interfaces hide the model's default tendencies, leaving users unaware that their ``original'' ideas may cluster with thousands of similar outputs. Our first goal is to make algorithmic consensus perceptible as a concrete object users can observe and reason about.

\textbf{Enable deliberate divergence with transparent control.} Visualization alone is insufficient---users need mechanisms to act on what they see. Drawing on Rafner et al.'s emphasis on process transparency~\cite{Rafner01122025}, we expose both the consensus landscape and the parameters that govern divergence strength, allowing users to make informed choices about how much novelty to pursue.

\textbf{Maintain coherence while diverging.} Increasing distance from consensus risks generating incoherent or low-quality outputs. We balance novelty with readability through fluency constraints and targeted penalties, ensuring that divergent outputs remain useful rather than merely different.

\subsection{Model Architecture}

SRT employs a dual-model contrastive decoding architecture using two instruction-tuned causal language models: Qwen2.5-7B-Instruct (strong model) and Qwen2.5-1.5B-Instruct (weak model) \cite{qwen2025qwen25technicalreport}. Both models support optional 4-bit NF4 quantization via \texttt{bitsandbytes} when CUDA is available. Since the two models use different tokenizers with vocabulary sizes $|V_s|$ and $|V_w|$, a pre-computed vocabulary alignment mapping $\mathbf{m} \in \mathbb{Z}^{|V_s|}$ matches token strings between models. For each strong-model token index~$i$, the mapping is $m_i = j$ if the token string matches weak index~$j$, and $m_i = {-}1$ otherwise. Typical vocabulary overlap is 70--80\%.

Semantic similarity throughout the system is computed via \\\texttt{all-MiniLM-L6-v2}, which maps variable-length text to 384-dimensional $L_2$-normalized embeddings, with cosine distance as the divergence metric:
\begin{equation}
d(\mathbf{u}, \mathbf{v}) = 1 - \mathbf{u}^\top \mathbf{v}
\label{eq:cosine_distance}
\end{equation}
where $\|\mathbf{u}\| = \|\mathbf{v}\| = 1$.

\subsection{User Workflow}

SRT transforms single-shot generation into a multi-stage interaction: (1)~Users enter prompts and select a task mode (Creative, Technical, or Brainstorm). (2)~The Hivemind Detector generates 12 samples to estimate the consensus distribution. (3)~The Semantic Radar visualizes consensus as a ``Yellow Zone'' on a 2D map. (4)~Users adjust a repulsion slider ($\lambda$) and generate baseline versus repulsed outputs side-by-side, each annotated with its cosine distance from the consensus centroid. (5)~Users compare results and iterate with different settings.

\subsection{Hivemind Detector}

Given a prompt $p$ and task mode $\mathcal{M}$, the Hivemind Detector generates $K = 12$ samples using nucleus sampling \cite{holtzman2019curious} (temperature $\tau_{\text{temp}}$, top-$p$ threshold $p_{\text{nucleus}}$, maximum 128 tokens). We chose $K = 12$ as it provides sufficient samples for UMAP \cite{McInnes2018} projection (which requires $K' \geq 6$ unique samples) while remaining computationally tractable. After deduplication---removing exact string duplicates---$K' \leq K$ unique samples remain.

\textbf{Centroid and Modal Sample.} Let $\{s_1, \ldots, s_{K'}\}$ denote unique samples and $\{\mathbf{e}_1, \ldots, \mathbf{e}_{K'}\}$ their $L_2$-normalized embeddings. The consensus centroid is:
\begin{equation}
\mathbf{c} = \frac{\textstyle\sum_{i=1}^{K'} \mathbf{e}_i}{\left\|\textstyle\sum_{i=1}^{K'} \mathbf{e}_i\right\|}
\label{eq:centroid}
\end{equation}
The modal sample minimizes cosine distance to~$\mathbf{c}$:
\begin{equation}
i^* = \underset{i}{\arg\min}\; \bigl(1 - \mathbf{e}_i^\top \mathbf{c}\bigr)
\label{eq:modal_sample}
\end{equation}
Sample $s_{i^*}$ represents ``what the model most wants to say.'' The detector also extracts \textit{negative concepts}---specific phrases characterizing consensus patterns.

\subsection{Negative Concept Extraction}
Negative concepts are consensus-characterizing phrases that the Repulsion
Engine penalizes during generation. Extraction differs by task mode to reflect
a fundamental difference in how consensus manifests across modes. In Creative
mode, consensus takes the form of prompt-specific narrative clichés that vary
with the topic (e.g., ``AI uprising'' for science fiction, ``dystopian
society'' for speculative fiction), so phrases must be extracted dynamically
from the sampled outputs. In Technical and Brainstorming modes, consensus
manifests instead as prompt-invariant surface-form boilerplate---stylistic
hedging and marketing formulae that appear regardless of the specific task
content---making a curated list both sufficient and more reliable than
dynamic extraction. This design choice is validated empirically: our ablation
study (Section~\ref{sec:ablation}) shows that phrase penalties reduce
consensus phrase frequency by 0.38 with near-zero coherence cost
($\Delta\text{Perplexity} = -89$ relative to a baseline of 4.5), confirming
that the mode-specific extraction strategy successfully targets surface-form
consensus without disrupting fluency.
 
\textbf{Creative Mode} uses YAKE~\cite{campos2020yake} keyword extraction
with maximum $n$-gram length $n_{\max} = 4$ and candidate pool size
$k_{\text{top}} = 120$, applied to concatenated sample text. Candidates are
filtered through five stages:
\begin{enumerate}
    \item \textit{Phrase-only filtering} requires $\geq 2$ alphabetic words
    per phrase.
    \item \textit{Generic content filtering} removes phrases where half or
    more content words belong to a predefined generic set (e.g., ``time,''
    ``people,'' ``things,'' ``way'') or to standard English stopwords.
    \item \textit{Cross-sample consensus filtering} retains only phrases
    appearing in $\geq 2$ of the $K'$ samples, ensuring they represent
    genuine consensus patterns rather than one-off expressions.
    \item \textit{Prompt exclusion} discards phrases whose words are entirely
    contained in the user's original prompt.
    \item The remaining phrases are ranked by
    $(\text{frequency}, -\text{YAKE score})$ and the top 12 are selected.
\end{enumerate}
 
\textbf{Technical Mode} uses a curated list of stylistic boilerplate
phrases (e.g., ``let's break down,'' ``let's delve into,'' ``in summary,''
``as you can see'') and returns only those observed in $\geq 1$ sample, up
to a maximum of 12 concepts. These phrases were selected based on documented
patterns of LLM-generated hedging and transitional language that signal
AI-produced text without contributing informational content~\cite{zhang2025}.
 
\textbf{Brainstorm Mode} uses a curated list of marketing frames (e.g.,
``revolutionary solution,'' ``game changer,'' ``cutting edge,'' ``think
outside the box'') as a primary source, selected to target the
over-representation of promotional framing in LLM brainstorming
outputs~\cite{jiang2025}. If fewer than 12 concepts are found, the Creative
mode YAKE extraction pipeline fills the remainder.

\subsection{Semantic Radar Visualization}
\label{sec:semantic_radar}

The Semantic Radar transforms high-dimensional embedding space into an interactive 2D landscape that makes consensus visible as a spatial object. We construct a matrix $\mathbf{E} \in \mathbb{R}^{(K'+1) \times 384}$ containing sample embeddings and the centroid (if a user draft is provided, its embedding is appended). When sufficient unique samples exist ($K' \geq 6$), we apply UMAP~\cite{mcinnes2018umap} with $\texttt{n\_neighbors} = \min(10, K'{-}1)$, $\texttt{min\_dist} = 0.15$, cosine metric, and fixed $\texttt{random\_state} = 42$. For smaller sets ($K' < 6$), we fall back to PCA with 2 components. Drawing on techniques from DataMap~\cite{ge2025datamapportableapplicationvisualizing} and latent space visualization research~\cite{liu2016visualizing}, we repurpose dimensionality reduction for a human-centered goal: making algorithmic consensus visible as a spatial object users can navigate around.

The signature ``Yellow Zone'' emerges from Gaussian kernel density estimation (KDE) over the projected sample positions. Let $\mathbf{X} \in \mathbb{R}^{K' \times 2}$ denote projected sample coordinates. The density estimate is:
\begin{equation}
\hat{f}(\mathbf{x}) = \frac{1}{K'} \sum_{i=1}^{K'} \mathcal{N}\!\bigl(\mathbf{x};\, \mathbf{X}[i],\, \mathbf{\Sigma}\bigr)
\label{eq:kde}
\end{equation}
where $\mathbf{\Sigma}$ is estimated via Scott's rule. We evaluate $\hat{f}$ on a $120 \times 120$ grid spanning the data range $\pm 0.8$ margin, and render it as a semi-transparent contour plot (opacity~$= 0.35$), where darker regions indicate higher consensus density. Sample points are displayed as scatter markers with hover text showing full sample content, the centroid is marked distinctly showing the modal sample, and an optional ``you'' marker shows the user's draft position if provided.

The spatial metaphor (distance~$=$~semantic distance) relies on the underlying embedding model, where relative positions in the 2D projection preserve relative semantic distances from the original high-dimensional space. Key design principles include progressive disclosure (overview to detail), personal positioning (users overlay drafts to see their location relative to consensus), and transparent representation (the centroid shows the modal sample text).

\subsection{Repulsion Engine}
\label{sec:repulsion_engine}

The Repulsion Engine provides a slider (``Baseline'' to ``Strong'') mapped to parameter $\lambda \in [0, 1.5]$ governing semantic distance from consensus. The engine combines three complementary mechanisms:

\textbf{Contrastive Decoding.} Inspired by~\cite{li2023contrastive} and \cite{liu2021dexpertsdecodingtimecontrolledtext}, at each generation step~$t$, we compute logits from both models and apply:
\begin{equation}
\boldsymbol{\ell}_{\text{ct}}^{(t)} = \boldsymbol{\ell}_s^{(t)} - \lambda \cdot \boldsymbol{\ell}_{w}^{(t)}
\label{eq:contrastive}
\end{equation}
where $\boldsymbol{\ell}_s^{(t)} \in \mathbb{R}^{|V_s|}$ are strong model logits and $\boldsymbol{\ell}_{w}^{(t)}$ are weak model logits aligned to the strong vocabulary via mapping~$\mathbf{m}$:
\begin{equation}
\ell_{w}^{(t)}[i] = 
\begin{cases}
\ell_{w,\text{raw}}^{(t)}[m_i] & \text{if } m_i \geq 0 \\
0 & \text{otherwise}
\end{cases}
\label{eq:alignment}
\end{equation}
This amplifies the strong model's distinctive capabilities while suppressing patterns both models share---effectively penalizing ``consensus'' predictions. Higher $\lambda$ produces greater suppression of shared patterns and thus greater divergence.

\textbf{Phrase-Level Penalties.} Building on unlikelihood training~\cite{welleck2019neuraltextgenerationunlikelihood}, the engine applies a direct logit penalty $\beta$ to all tokens belonging to negative concept phrases extracted by the Hivemind Detector:
\begin{equation}
\ell_{\text{ct}}^{(t)}[i] \leftarrow \ell_{\text{ct}}^{(t)}[i] - \beta, \;\; \forall\, i \in \mathcal{T}_{\text{neg}}
\label{eq:phrase_penalty}
\end{equation}
where $\mathcal{T}_{\text{neg}} = \bigcup_{\text{ph} \in \mathcal{N}} \text{tok}(\text{ph})$ is the union of all tokens appearing in negative concept phrases, and $\beta = 4.0$ across all modes. Because the penalty operates at the token level, it suppresses not only exact consensus phrases but also their constituent words in other contexts---a deliberate trade-off favoring broader consensus avoidance over surgical precision.

\textbf{Fluency and Diversity Controls.} Three mechanisms maintain output quality during divergence:

\noindent\textit{(1)~Fluency floor.} Tokens whose strong-model probability falls below threshold~$\tau$ are masked:
\begin{equation}
\ell_{\text{ct}}^{(t)}[i] \leftarrow {-}\infty \;\;\text{if}\;\; p_s^{(t)}(i) < \tau
\label{eq:fluency_floor}
\end{equation}
where $p_s^{(t)}(i) = \text{softmax}(\boldsymbol{\ell}_s^{(t)})[i]$, with $\tau = 0.003$ for Creative, $0.010$ for Technical, and $0.002$ for Brainstorm.

\noindent\textit{(2)~Repetition penalty.} For each previously generated token $i \in \mathbf{y}_{<t}$:
\begin{equation}
\ell_{\text{ct}}^{(t)}[i] \leftarrow 
\begin{cases}
\ell_{\text{ct}}^{(t)}[i] / \rho & \text{if } \ell_{\text{ct}}^{(t)}[i] > 0 \\
\ell_{\text{ct}}^{(t)}[i] \cdot \rho & \text{otherwise}
\end{cases}
\label{eq:repetition_penalty}
\end{equation}
with $\rho = 1.06$ for Creative/Brainstorm and $\rho = 1.12$ for Technical.

\noindent\textit{(3)~$N$-gram blocking.} We set $\ell_{\text{ct}}^{(t)}[i] \leftarrow {-}\infty$ for any token~$i$ that would complete a repeated $n$-gram ($n = 4$ for Creative/Brainstorm, $n = 6$ for Technical).

Additionally, tokens containing non-ASCII alphabetic characters (excluding a mathematical symbol whitelist $\Lambda = \{\lambda, \gamma, \mu, \sigma,  \pi, \theta, \alpha, \\ \beta, \delta, \epsilon, \kappa, \rho, \nu, \tau\}$) are blocked throughout generation. After temperature scaling and top-$p$ filtering, the final token is sampled from the resulting distribution. If the softmax produces NaN or zero sum, the engine falls back to strong-model-only sampling with the same constraints. In Creative mode, an optional polish pass uses greedy decoding (max 240 tokens) to fix spelling and grammar without reintroducing penalized phrases.

\textbf{Practical Interpretation of $\lambda$.} The slider provides predictable control: higher values consistently produce greater semantic distance from consensus. At $\lambda = 0$ (baseline), the system generates using the strong model alone. At $\lambda = 0.6$ (SRT-Mild), the weak model's contribution is moderately subtracted, producing stylistic variation while staying relatively close to default patterns. At $\lambda = 1.2$ (SRT-Strong), the subtraction is substantial, yielding outputs with markedly different framing, vocabulary, and structure. Users see both baseline and repulsed outputs side-by-side, each annotated with its cosine distance from the centroid, enabling direct comparison of the divergence trade-off.

Algorithm~\ref{alg:srt} summarizes the core Semantic Repulsion Technique (SRT) pipeline. 
Given a user prompt, SRT first samples the model's default response distribution to estimate 
the consensus region, extracts consensus-characterizing phrases, and then performs 
repulsion-based contrastive decoding to generate an output that moves away from common 
model patterns while preserving fluency.

\subsection{Task Modes}

Mode-specific defaults reflect varying requirements for accuracy versus exploration. Table~\ref{tab:hyperparameters} summarizes the full configuration.

\begin{table}[t]
\caption{Hyperparameters by task mode. Interface defaults 
are $\lambda = 1.2$ (Creative), $\lambda = 0.6$ 
(Technical), and $\lambda = 1.5$ (Brainstorm). For the 
controlled evaluation, $\lambda = 0.6$ (SRT-Mild) and 
$\lambda = 1.2$ (SRT-Strong) were applied uniformly 
across all modes.}
\label{tab:hyperparameters}
\centering
\small
\begin{tabular}{lccc}
\toprule
\textbf{Parameter} & \textbf{Cre.} & \textbf{Tech.} & \textbf{Brain.} \\
\midrule
$\lambda$ (repulsion)       & 1.2 & 0.6 & 1.5 \\
$\tau$ (fluency floor)      & 0.003 & 0.010 & 0.002 \\
$\beta$ (phrase penalty)    & 4.0 & 4.0 & 4.0 \\
$\rho$ (rep.\ penalty)      & 1.06 & 1.12 & 1.06 \\
$n$ (ngram block)            & 4 & 6 & 4 \\
$L_{\text{gen}}$ (tokens)   & 320 & 220 & 260 \\
Baseline dec.                & Samp. & Greedy & Samp. \\
Polish pass                  & Opt. & No & No \\
\bottomrule
\end{tabular}
\end{table}

Creative mode ($\lambda = 1.2$) targets narrative clich\'{e}s using YAKE-extracted consensus phrases. Technical mode ($\lambda = 0.6$) uses  a higher fluency floor ($\tau = 0.010$) and stricter repetition constraints ($\rho = 1.12$, $n = 6$) to preserve factual accuracy while targeting stylistic boilerplate. Brainstorm mode ($\lambda = 1.5$) targets marketing frames and allows the most aggressive exploration with the lowest fluency floor ($\tau = 0.002$).

\section{System Evaluation}
\label{sec:evaluation}

We validate SRT through three complementary computational assessments:
(1)~a controlled comparison establishing that SRT produces measurably more
divergent outputs than baseline sampling strategies; (2)~a mechanism ablation
isolating the causal contribution of each SRT component; and (3)~an
operating-range analysis characterizing the divergence--relevance--coherence
trade-off across the full $\lambda$ range.  Together these address the core
technical claims: that SRT diverges from consensus, that each mechanism serves
a distinct and necessary role, and that the system remains prompt-relevant
within its intended operating range.

\subsection{Computational Assessment}
\label{sec:exp1}

\textbf{Setup.}
We conducted a controlled comparison across 30 prompts (10 per task mode),
comparing five systems: \textbf{Baseline-Pure} (nucleus
sampling~\cite{holtzman2019curious}, temp~$= 1.0$, top-$p = 0.9$),
\textbf{Baseline-HighTemp} (temp~$= 1.5$), \textbf{Baseline-Beam}
($\texttt{num\_beams} = 5$), \textbf{SRT-Mild} ($\lambda = 0.6$), and
\textbf{SRT-Strong} ($\lambda = 1.2$). All systems used
Qwen2.5-7B-Instruct; SRT variants additionally employed contrastive decoding
with Qwen2.5-1.5B-Instruct as the weak model. Values of $\lambda$ were
applied uniformly across all modes to isolate the effect of contrastive
repulsion. For each prompt, the Hivemind Detector established consensus by
generating 12 samples, computing 384-dimensional embeddings via
\texttt{all-MiniLM-L6-v2}, calculating the consensus centroid, and
extracting negative concepts via YAKE~\cite{campos2020yake}. We then
generated 10 outputs per system per prompt (1,500 total).

\textbf{Metrics.}
We measured three complementary aspects. \textit{Originality} (distance from
consensus centroid):
\begin{equation}
D_{\text{mode}}(g) = 1 - \mathbf{e}_g^\top \mathbf{c}
\label{eq:originality}
\end{equation}
Higher values indicate greater divergence from defaults.
\textit{Diversity} (intra-system): mean pairwise cosine distance between all
outputs from the same system for the same prompt.
\textit{Consensus Phrase Frequency}: count of negative concepts appearing in
each output; lower values indicate greater consensus suppression.
We note that embedding-based originality measures semantic distance from the
consensus but cannot distinguish diversity of expression from diversity of
ideas; this limitation is acknowledged in Section~\ref{sec:ref}.

\begin{figure*}[t]
\centering
\includegraphics[width=\textwidth]{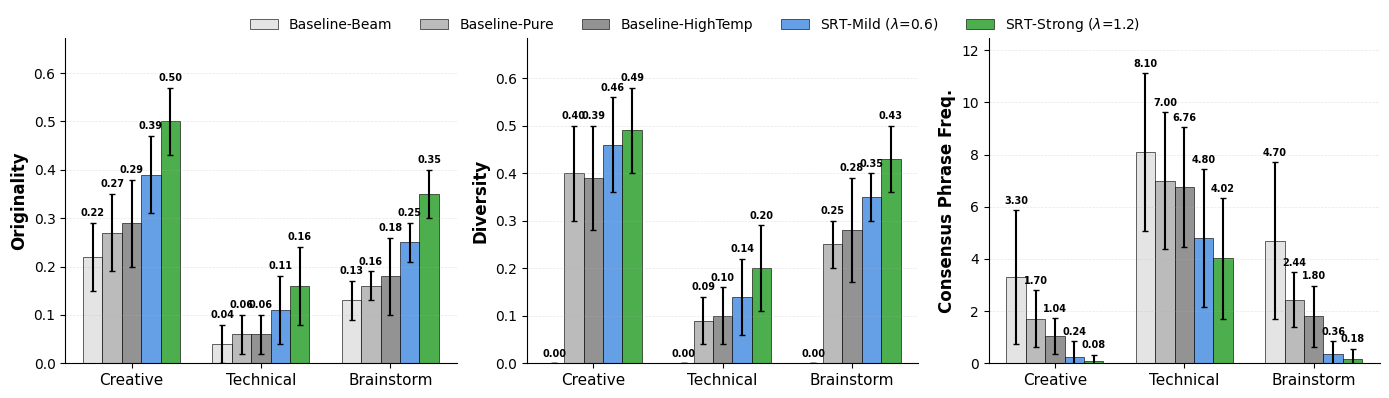}
\caption{\textbf{Computational evaluation results across task modes.}
\textbf{(a)}~Originality: distance from consensus centroid.
\textbf{(b)}~Diversity: intra-system variance.
\textbf{(c)}~Consensus Phrase Frequency: count of consensus phrases.
SRT-Strong (green) achieves 85\%, 167\%, and 119\% originality improvements
over Baseline-Pure (gray) in Creative, Technical, and Brainstorm modes.
SRT-Mild (blue) shows intermediate performance, demonstrating controllable
divergence via $\lambda$. Error bars show standard deviation
(10~samples per condition).}
\label{fig:computational}
\end{figure*}

\textbf{Results.}
Results strongly support SRT's divergence capabilities across all three task
modes (Figure~\ref{fig:computational}).

\textit{Originality.} SRT-Strong achieved substantially higher divergence than
all baselines across every mode. In Creative mode, SRT-Strong outputs were
nearly twice as distant from the centroid as Baseline-Pure ($0.50$
vs.\ $0.27$)---an 85\% improvement. Technical mode showed the largest
relative gain ($+167\%$), and Brainstorm mode showed $+119\%$ improvement.
SRT-Mild showed intermediate performance in all modes, demonstrating
controllable divergence via $\lambda$. Baseline-Beam exhibited the lowest
divergence across all modes, confirming that probability maximisation produces
highly consensual outputs. Baseline-HighTemp showed high variance without
consistently higher originality, suggesting unfocused divergence.

\textit{Diversity.} SRT maintained or improved intra-system diversity compared
to sampling baselines, demonstrating broader semantic exploration rather than
convergence to a new narrow peak. In Creative mode, SRT-Strong achieved
diversity of $0.49$ compared to $0.40$ for Baseline-Pure while simultaneously
achieving higher originality---evidence against simple mode substitution.
Baseline-Beam showed zero diversity by design.

\textit{Consensus Phrase Frequency.} SRT-Strong achieved near-zero counts in
Creative mode ($0.08$)---a 95\% reduction versus Baseline-Pure ($1.70$) and
98\% versus Baseline-Beam ($3.30$). Brainstorm mode showed 93\% reduction.
Technical mode exhibited higher absolute counts due to boilerplate detection
methodology, but SRT-Strong still reduced consensus phrases by 43\% versus
Baseline-Pure.

\textit{Mode-Specific Effectiveness.} Creative and Brainstorm modes, which
tolerate greater semantic exploration, showed larger originality gains.
Technical mode, configured with a higher fluency floor ($\tau = 0.010$) and
stricter repetition constraints to preserve factual accuracy, demonstrated
meaningful divergence while maintaining coherence---evidence that the system
successfully navigates the novelty--accuracy trade-off.

\begin{figure}[t]
\centering
\includegraphics[width=\columnwidth]{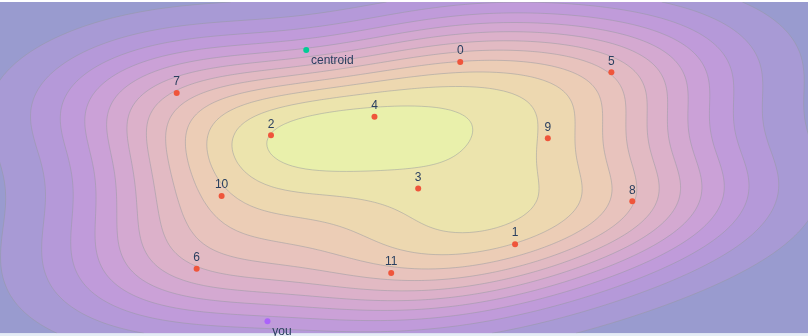}
\caption{The Semantic Radar's Yellow Zone visualization. Gaussian KDE renders
consensus density as a heatmap over UMAP-projected sample embeddings; darker
regions indicate higher response clustering. The centroid (marked) represents
the modal sample. Users can overlay their own drafts to see their position
relative to consensus.}
\label{fig:YZ}
\end{figure}

\subsection{Mechanism Ablation}
\label{sec:ablation}

The computational assessment establishes that SRT diverges from consensus, but
does not indicate which component drives which outcome. We address this through
a 2$^3$ factorial ablation across 1,200 generated outputs (30 prompts
$\times$ 8 conditions $\times$ 5~samples), toggling each of the three SRT
mechanisms---Contrastive Decoding (CD), Phrase Penalties (PP), and Fluency
Controls (FL)---independently. In addition to the three original metrics, we
report two new metrics that directly concerns with
prompt-relevance and objective coherence: \textit{Relevance}, the cosine
similarity between the prompt embedding and the output embedding (higher =
more on-topic); and \textit{Coherence}, the negative mean token
log-likelihood under the strong model (higher = more fluent), also expressed
as \textit{Perplexity} ($\exp(-\text{NLL})$; lower = more fluent).

\begin{figure}[t]
\centering
\includegraphics[width=\columnwidth]{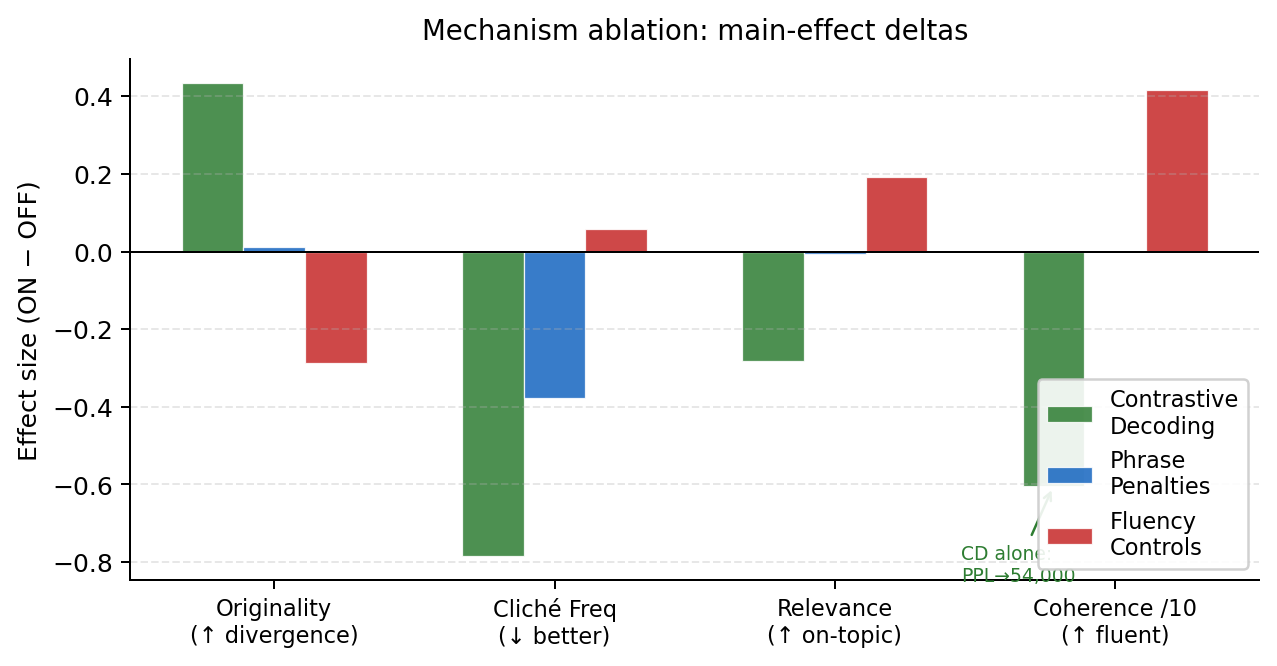}
\caption{Mechanism ablation main-effect deltas. Each cluster shows the effect
of enabling one mechanism (ON $-$ OFF), averaged across the other mechanisms'
states. Coherence column is divided by~10 for display. Contrastive Decoding
drives divergence but is unsafe alone (PPL\,${\to}\,54{,}000$); Phrase
Penalties provide surgical clich\'{e} suppression at near-zero coherence cost;
Fluency Controls are the essential stabilizer.}
\label{fig:ablation}
\end{figure}

\textbf{Results.}
Table~\ref{tab:ablation} reports the main-effect delta for each mechanism---the
mean difference in each metric when the mechanism is ON versus OFF, averaged
across all conditions in which the other two mechanisms vary. Each mechanism
serves a distinct and necessary role (Figure~\ref{fig:ablation}).

\textit{Contrastive Decoding} is the primary driver of divergence
($\Delta\text{Originality} = {+}0.43$) and the dominant source of clich\'{e}
suppression ($\Delta\text{Clich\'{e}} = {-}0.79$). Used in isolation,
however, CD produces catastrophically incoherent output---perplexity rises
from a baseline of~4.5 to ${\sim}54{,}000$ (C1\_CD only condition)---because
the contrastive subtraction suppresses likely tokens without any floor on
impossible ones. This confirms that CD is a powerful but unsafe mechanism
when deployed alone.

\textit{Phrase Penalties} provide surgical clich\'{e} suppression
($\Delta\text{Clich\'{e}} = {-}0.38$) with near-zero impact on all other
metrics ($\Delta\text{Perplexity} = {-}89$, negligible relative to the
baseline of~4.5; $\Delta\text{Originality} = {+}0.01$;
$\Delta\text{Relevance} = {-}0.007$). This cleanly justifies the mode-specific
extraction design: phrase penalties target surface-form consensus without
interfering with the embedding-space repulsion signal.

\textit{Fluency Controls} are the essential stabilizer: they reduce perplexity
by $53{,}907$ points ($\Delta\text{Coherence} = {+}4.16$), recovering from
the CD-induced incoherence while accepting a modest originality cost
($\Delta\text{Originality} = {-}0.29$). The positive relevance effect
($\Delta\text{Relevance} = {+}0.19$) shows that fluency controls also restore
on-topic quality.

The full SRT configuration (C7) balances these contributions: in Creative mode
it achieves originality~$0.48$ versus baseline~$0.29$ ($+65\%$), reduces
clich\'{e} frequency from~$1.38$ to~$0.04$ ($-97\%$), and maintains
perplexity at~$71$---substantially higher than the baseline of~$5.9$ but five
orders of magnitude lower than CD alone. All Wilcoxon signed-rank comparisons
of C7 versus C0 (baseline) are $p < 0.001$ across all metrics and modes,
confirming that the observed effects are not attributable to sampling variance.

\begin{table}[t]
\centering
\small
\caption{Mechanism ablation: main-effect deltas (ON $-$ OFF), averaged across
all conditions and modes ($N = 150$ per mechanism state). $\Delta$Coherence
is the change in negative NLL (higher = more fluent).}
\label{tab:ablation}
\setlength{\tabcolsep}{4.5pt}
\begin{tabular}{lrrrrr}
\toprule
\textbf{Mechanism} & $\Delta$\textbf{Orig.} & $\Delta$\textbf{Div.} &
$\Delta$\textbf{Cliché} & $\Delta$\textbf{Rel.} & $\Delta$\textbf{Coh.} \\
\midrule
Contrastive Decoding & $+0.434$ & $+0.149$ & $-0.785$ & $-0.283$ & $-6.03$ \\
Phrase Penalties     & $+0.012$ & $+0.002$ & $-0.378$ & $-0.007$ & $-0.02$ \\
Fluency Controls     & $-0.287$ & $-0.028$ & $+0.058$ & $+0.191$ & $+4.16$ \\
\bottomrule
\end{tabular}
\end{table}

\subsection{Operating Range Analysis}
\label{sec:lambda}

A key open question is how far repulsion strength $\lambda$ can be increased
before outputs lose prompt relevance.
We address this through a $\lambda$-sweep study across nine values
$\lambda \in \{0.0, 0.6, 1.2, 1.8, 2.4\}$ with phrase penalties and fluency
controls held fixed, generating 750 outputs (30 prompts $\times$ 5 $\lambda$
values $\times$ 5 samples). The same five metrics as the ablation study are
reported.

\begin{figure*}[t]
\centering
\includegraphics[width=\textwidth]{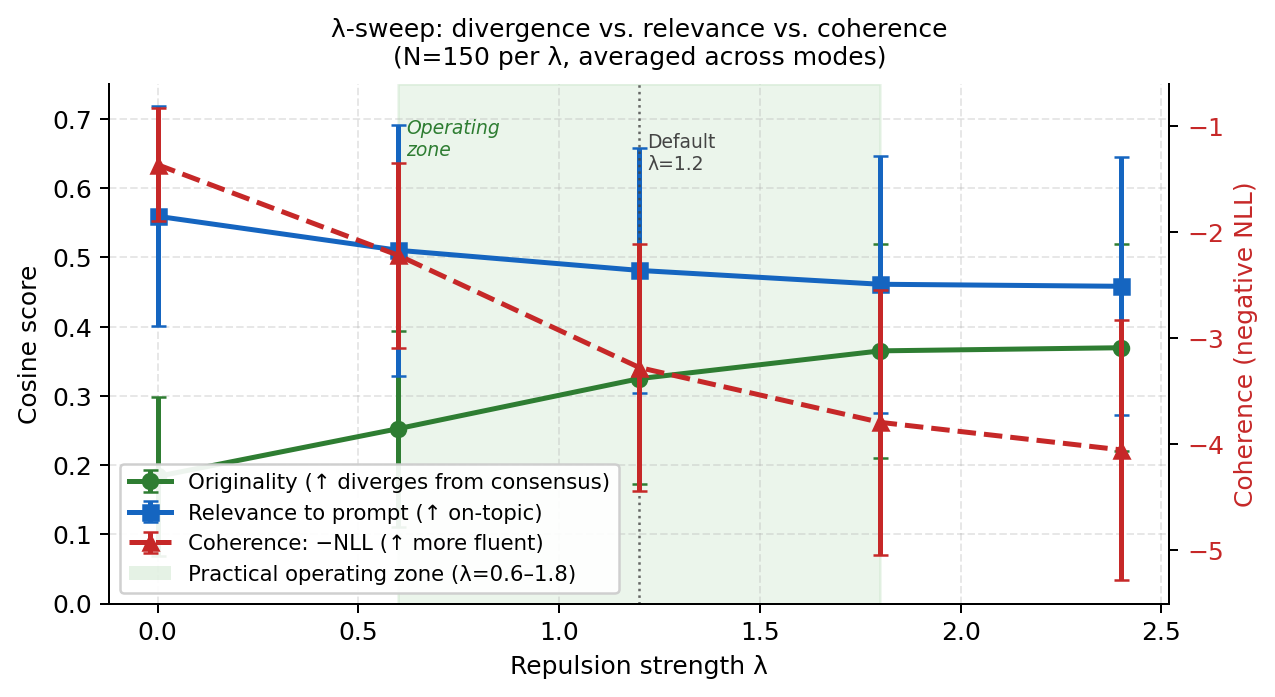}
\caption{$\lambda$-sweep: originality (green), relevance (blue), and
coherence/$-$NLL (red, right axis) as a function of repulsion strength.
Shaded region indicates the practical operating zone ($\lambda = 0.6$--$1.8$)
where originality gains are large and relevance remains above 82\% of
baseline. Error bars show SD ($N = 150$ per point).}
\label{fig:lambda}
\end{figure*}

\textbf{Results.}
Table~\ref{tab:lambda} and Figure~\ref{fig:lambda} show the divergence--%
relevance--coherence trade-off across the operating range.

Originality increases monotonically from $0.18$ at $\lambda = 0$ to $0.36$
at $\lambda = 1.8$ ($+99\%$), then plateaus: the gain from $\lambda = 1.8$
to $\lambda = 2.4$ is only $+0.005$ ($+1.4\%$). Relevance declines gradually,
falling from $0.56$ at baseline to $0.46$ at $\lambda = 2.4$---a maximum drop
of only $18.1\%$, indicating that outputs remain substantially on-topic across
the full range tested. Computational coherence degrades more sharply
(perplexity $4.5 \to 92.2$), though the user study shows that perceived
coherence is \textit{higher} for SRT-Strong than for the baseline
($M = 3.92$ vs.\ $3.48$, $p = .006$), suggesting that self-perplexity
overestimates the readability cost of divergence for human readers.

The diminishing returns in originality beyond $\lambda = 1.8$, combined with
continued declines in relevance and coherence, establish $\lambda \approx
1.2$--$1.8$ as the practical operating range. The interface defaults of
$\lambda = 1.2$ (Creative), $\lambda = 0.6$ (Technical), and $\lambda = 1.5$
(Brainstorm) all fall within this empirically validated zone. Technical mode
shows notably greater robustness: relevance drops only $8\%$ from baseline
even at $\lambda = 2.4$ ($0.65$ vs.\ $0.71$), validating its lower default
configuration. All Wilcoxon tests comparing each $\lambda > 0$ to $\lambda = 0$
are $p < 0.0001$ across all metrics, confirming that even the smallest tested
step ($\lambda = 0.6$) produces statistically reliable divergence.

\begin{table}[t]
\centering
\small
\caption{$\lambda$-sweep summary averaged across all modes
($N = 150$ per $\lambda$, mean values shown).}
\label{tab:lambda}
\setlength{\tabcolsep}{4pt}
\begin{tabular}{lrrrrr}
\toprule
$\lambda$ & \textbf{Orig.} & \textbf{Rel.} & \textbf{Cliché} &
\textbf{Coh.} & \textbf{PPL} \\
\midrule
0.0 & 0.184 & 0.559 & 0.513 & $-1.37$ & 4.5  \\
0.6 & 0.253 & 0.510 & 0.113 & $-2.22$ & 12.5 \\
1.2 & 0.325 & 0.481 & 0.073 & $-3.28$ & 42.8 \\
1.8 & 0.365 & 0.461 & 0.047 & $-3.80$ & 72.2 \\
2.4 & 0.370 & 0.458 & 0.020 & $-4.06$ & 92.2 \\
\bottomrule
\end{tabular}
\end{table}

\subsection{User Study}
 
To evaluate whether computational divergence translates into perceived utility
and adoption intent, we conducted a within-subjects exploratory study with
$N=16$ participants (ages 22--48; 4 female, 12 male) who regularly use AI for
creative tasks (Table~\ref{tab:demographics}). We treat this as a preliminary
investigation complementing the computational validation in
Section~\ref{sec:evaluation}; findings should be interpreted as suggestive
rather than generalisable given the sample size.
 
\textbf{Procedure.} Participants completed three tasks: (1)~Creative Writing
(rewriting a sci-fi story premise), (2)~Technical Writing (explaining a topic
of interest), and (3)~Brainstorming (generating ideas). For each task, they
reviewed \textit{four} AI-generated responses---Baseline-Pure,
Baseline-HighTemp, SRT-Mild, SRT-Strong---yielding 12 response evaluations
per participant in total (3~tasks $\times$ 4~systems). Each system produced a
single output per task to enable controlled within-subjects comparison across
all four conditions simultaneously; iterative human--AI co-writing is an
important direction we leave to future work. After reviewing all responses,
participants rated each on 5-point Likert scales for originality, coherence,
creativity, and usefulness, indicated willingness to use, and provided open
feedback. Following all tasks, we presented the Yellow Zone visualization
(Figure~\ref{fig:YZ}), explained consensus detection, and asked participants
to select which system(s) they would most want to use for creative projects.
System preferences were elicited \textit{after} the Yellow Zone explanation,
which may have primed participants toward consensus-aware systems; these
results should therefore be interpreted alongside the blind Likert ratings.
 
\begin{table}[h]
\centering
\caption{Participant demographics ($N = 16$).}
\label{tab:demographics}
\small
\begin{tabular}{l l}
\hline
\textbf{Characteristic} & \textbf{Summary} \\
\hline
\textit{Age} & \\
\quad Range & 22--48 \\
\quad Mean (SD) & 28.1 (6.3) \\
\quad 18--24 & 2 (12.5\%) \\
\quad 25--30 & 11 (68.8\%) \\
\quad 31--50 & 3 (18.8\%) \\
\hline
\textit{Gender} & \\
\quad Male & 12 (75.0\%) \\
\quad Female & 4 (25.0\%) \\
\hline
\textit{AI Tool Usage Frequency} & \\
\quad Daily & 12 (75.0\%) \\
\quad Weekly & 3 (18.8\%) \\
\quad Rarely & 1 (6.3\%) \\
\hline
\textit{Creative Writing Frequency} & \\
\quad Daily & 8 (50.0\%) \\
\quad Weekly & 6 (37.5\%) \\
\quad Rarely & 2 (12.5\%) \\
\hline
\textit{AI Tools Used (multi-select)} & \\
\quad ChatGPT & 12 (75.0\%) \\
\quad Claude & 6 (37.5\%) \\
\quad Gemini & 5 (31.3\%) \\
\quad Other (DeepSeek, Copilot, etc.) & 6 (37.5\%) \\
\hline
\end{tabular}
\end{table}
 
\textbf{Analysis.} We used non-parametric tests appropriate for ordinal Likert
data and small samples. For each measure, we aggregated ratings across three
tasks per participant-system combination, then conducted Friedman tests to
detect overall differences, followed by pairwise comparisons where warranted.
 
\begin{figure*}[t]
\centering
\includegraphics[width=\textwidth]{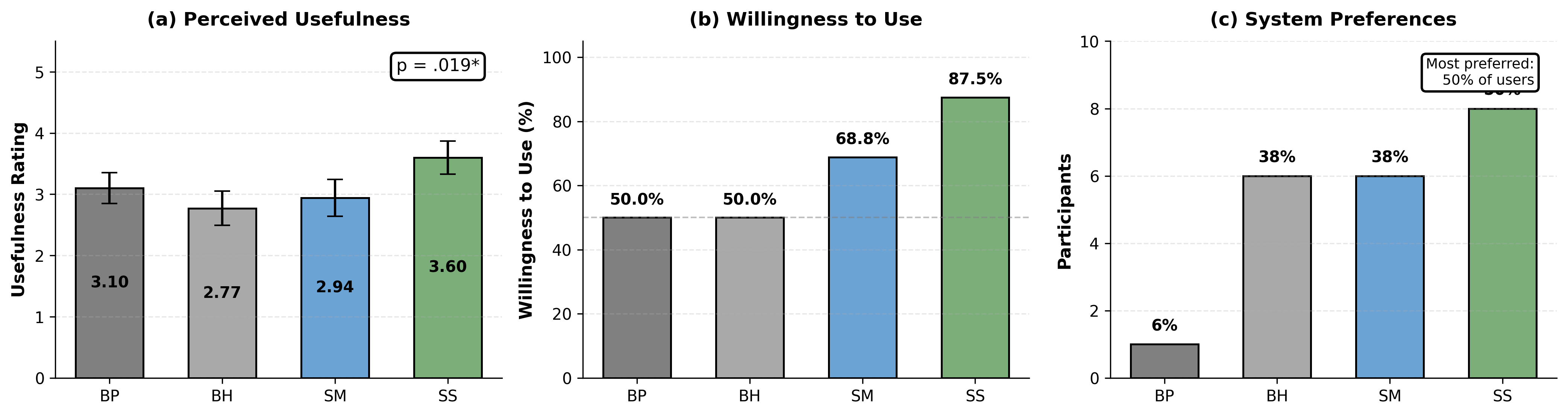}
\caption{User study results demonstrating practical utility and adoption
intent. \textbf{(a)}~Perceived usefulness ratings across systems (Friedman
$p=.019$). \textbf{(b)}~Willingness to use as a creative starting point,
shown as percentage of participants expressing willingness per task,
aggregated across all three tasks (Cochran's Q $p=.075$).
\textbf{(c)}~System preferences when given explicit choice; two participants
selected ``None.'' Error bars show 95\% confidence intervals. Systems:
BP=Baseline-Pure, BH=Baseline-HighTemp, SM=SRT-Mild, SS=SRT-Strong
($N=16$).}
\label{fig:user_results}
\end{figure*}

\textbf{Results.} Figure~\ref{fig:user_results} shows convergent evidence
across multiple utility dimensions.
 
\textit{Perceived Usefulness.} SRT-Strong received highest usefulness ratings
($M=3.60$), 16\% higher than Baseline-Pure ($M=3.10$). A Friedman test
revealed significant differences, $\chi^2(3, N=16)=9.99$, $p=.019$,
Kendall's $W=.208$ (small-to-medium effect). While the sample size is modest,
the effect size provides preliminary evidence of meaningful utility
differences.
 
\textit{Perceived Coherence.} SRT-Strong also received significantly higher
coherence ratings ($M=3.92$, $\mathit{Mdn}=4.00$) compared to Baseline-Pure
($M=3.48$, $\mathit{Mdn}=3.50$). A Friedman test revealed significant
differences, $\chi^2(3, N=16) = 12.47$, $p = .006$, Kendall's $W = .260$
(medium effect). This finding is notable because it suggests that semantic
divergence does not compromise---and may even enhance---perceived readability,
a result consistent with the positive originality--coherence correlations
observed below.
 
\textit{Willingness to Use.} Beyond ratings, behavioral intent showed a clear
pattern favoring consensus-aware systems. At the participant level, 87.5\% of
participants (14/16) indicated willingness to use SRT-Strong for at least one
task, compared to 50.0\% (8/16) for Baseline-Pure
(Figure~\ref{fig:user_results}b). Under a stricter threshold---willingness for
the majority of tasks (2+ of 3)---the gap widened further: 68.8\% (11/16)
for SRT-Strong versus only 18.8\% (3/16) for Baseline-Pure, a $3.7\times$
difference. A Cochran's Q test for repeated binary measures showed a trend
toward difference across systems, $Q(3) = 6.91$, $p = .075$. While not
reaching conventional significance, the consistency of preference across both
thresholds suggests genuine adoption intent.
 
\textit{System Preferences.} When asked to select which system(s) they would
most want to use (participants could select multiple), 50\% chose
SRT-Strong---the single most preferred option. Baseline-HighTemp and SRT-Mild
were each selected by 38\%, suggesting that stylistic variation is valued
whether achieved through higher temperature or mild repulsion. Grouping by
type, consensus-aware systems (SRT-Mild + SRT-Strong) received 67\% of all
selections (14/21) vs 33\% for baselines (7/21). Standard baseline sampling
was selected by only one participant (6\%), suggesting that users value some
form of divergence capability when given explicit choice.
 
\textit{Originality, Coherence, and Transparency.} Originality ratings
favoured SRT-Strong ($M=3.40$) over Baseline-Pure ($M=2.85$, $+19\%$),
though this difference approached but did not reach significance,
$\chi^2(3, N=16) = 7.34$, $p = .062$, $W = .153$. This pattern---significant
practical utility ($p=.019$) and coherence ($p=.006$) alongside marginal
perceived originality ($p=.062$)---suggests that users reliably detect the
\textit{usefulness} of divergent outputs without reliably detecting
\textit{originality} in blind evaluation. This is consistent with the
broader framing of SRT as a consensus-avoidance tool: awareness of consensus
may matter more than perceptual sensitivity to originality as such.
Critically, originality and coherence ratings were positively correlated
across all systems ($\rho = +.40$ to $+.67$), contradicting the expected
readability-novelty trade-off and providing indirect evidence that divergence
does not produce off-topic wandering. Participants rated SRT's transparency
features highly ($M=3.62$--$4.38/5$), with 68.8\% indicating they would
adopt such features in future AI tools.
 
\textit{Consensus Visualization.} When asked to interpret the Yellow Zone
before explanation, only 4/16 (25\%) correctly identified it as consensus
clustering; 6/16 (38\%) misinterpreted it as indicating optimal responses.
Rather than treating this as a study failure, we read it as a design finding:
users interpret visual density as quality rather than frequency. Future
consensus visualizations should use explicit labelling (e.g., ``common'' vs.\
``good'') and progressive disclosure to prevent this conflation---a design
implication independent of whether SRT's underlying divergence mechanism is
effective.

\textbf{Qualitative Results.}
We conducted thematic analysis~\cite{Braun01012006} on all open-ended responses, comprising per-response feedback (4 responses $\times$ 3 tasks) and 8 post-task reflection questions. Following Braun and Clarke's six-phase process, we familiarized ourselves with the full dataset, generated initial codes, searched for candidate themes, reviewed themes against coded extracts, and refined definitions iteratively. Response depth varied substantially: 11 participants provided detailed open-ended feedback, while 5 gave minimal or single-word responses on several items. Prevalence counts below report the number of participants contributing at least one substantive coded excerpt. Four themes were identified.

Before reporting themes, we note a key finding regarding
visualization interpretability. When asked to interpret the Yellow Zone before explanation, only 4/16 (25\%) correctly identified it as consensus clustering (e.g., P3: ``a place where similar concepts clustered''). Six participants (38\%) misinterpreted it as indicating optimal responses (e.g., P9:``most accurate responses'').

\textit{Theme 1: Coherence as Non-Negotiable Baseline} (13/16). The most consistent pattern was that participants wanted distinctive outputs but treated coherence as a non-negotiable precondition. P13 captured this directly: ``I need divergent writing but coherent---I don't want my essay to be same as others but I don't want it to be messy either.'' P12 stated: ``I would always choose coherence over creativity.'' When outputs lacked readability, participants rejected them regardless of novelty---P8 described one response as ``jargon with too much complex words, very unpleasant for the reader,'' while P14 dismissed another with ``What??'' Conversely, the desire for differentiation was strong: P3 stated ``I want my responses to be different than those commonly generated,'' and P12 asked ``Why would I write a story that 500 other people already did!'' This conditional framing contextualizes the quantitative coherence finding ($p = .006$): participants actively screened for readability, yet SRT-Strong scored highest, suggesting its fluency controls maintained the baseline participants demanded. Simultaneously, 8/16 expressed overcorrection fears---P14 warned ``If you max it out, the text would become trash to read. Not smooth at all,'' and P15 worried the result would be ``too different to my request.'' These fears persisted even though the quantitative data showed positive originality-coherence correlations ($\rho = +.40$ to $+.67$), suggesting participants hold intuitive mental models where novelty costs readability---an expectation that did not materialize within the tested
$\lambda$ range.

\textit{Theme 2: Strategic Rather Than Maximal Divergence} (13/16). Participants overwhelmingly preferred context-dependent use of divergence control. Eleven participants preferred context-dependent slider
usage; only 3/16 indicated they would maximize divergence,
with the remaining 2 preferring typical levels or finding
the control not yet useful. P8 stated: ``I would like to have a control of the text divergence and use it according to project.'' P5 described a multi-system workflow: ``probably a mix of both---start with System~B then D,'' and P10 similarly wanted to ``combine B and D.'' Evaluation criteria also shifted by task mode: in Technical writing, P12 noted ``I do not consider the creativity matter in this topic,'' while in Brainstorming, P8 valued ideas ``usable in the real world scenario'' and P15 praised responses that were ``creative-practical''---suggesting users apply different evaluative standards that align with the system's mode-specific configurations. Several participants conceptualized AI output as raw material for human editing rather than finished product: P16 explained wanting to ``see the most frequent combinations first'' and then ``revise to make the writing more creative later,'' while P8 stated a preference for editing ``the prewritten text by myself.''

\textit{Theme 3: Consensus Awareness} (10/16). A
majority reported that the study changed how they think
about AI-generated content. P2 stated: ``I never thought
about consensus before, now I'll look at responses from
the lens of consensus.'' P7 noted that ``default settings
clearly push people toward narrow conceptual and stylistic
basins,'' and P16 reported: ``I will pay more attention to
the homogenization content provided by AI.'' Among the
6/16 who reported no shift, reasons included pre-existing
awareness (P1, who already recognized homogenization as
``regression to the mean,'' saw no new shift) and general
skepticism (P11). When asked directly whether AI
homogenization is a real concern, 10/16 agreed, though
depth varied: P12 framed it as ``a natural outcome'' rather
than a design problem, while P4 argued effective prompting
overcomes convergence.

\textit{Theme 4: Divergence Skeptics} (3/16). Not all participants embraced automated divergence. P1, a daily creative writer who rarely uses AI, rejected all systems (``All generated absolutely awful content'') and reported no change in thinking, representing a consistent stance that AI-generated text is unsuitable for creative work regardless of divergence settings. P16---the only participant to prefer System~A (Standard AI)---wanted consensus visibility as \textit{information} rather than as a trigger for automated divergence: ``Choosing the creative AI systems can make my writing less human-like and I might put into more efforts.'' P2 raised a saturation concern: ``if I depend too much on AI for consensus avoidance\ldots wouldn't we reach saturation?'' These disconfirming cases indicate that consensus-aware generation is not universally desired; some users prefer manual editing over automated divergence, and others question whether systematic consensus-avoidance introduces its own limitations. We note that as system designers, our analytical lens may orient toward divergence-positive interpretations; we actively sought and report these disconfirming cases to counterbalance that tendency.

\section{Discussion and Conclusion}
\label{sec:ref}
SRT demonstrates that semantic divergence need not sacrifice
coherence---a finding supported by both direct ratings and their relationship. SRT-Strong achieved higher ratings for coherence ($M = 3.92$, $p = .006$, $W = .260$) and usefulness ($M = 3.60$, $p = .019$, $W = .208$), with the highest originality ratings ($M = 3.40$, $p = .062$) among all systems. Originality and coherence ratings were positively correlated across all systems ($\rho = +.40$ to $+.67$), indicating that participants who perceived outputs as more original also tended to rate them as more coherent. This positive relationship also provides indirect evidence against mere semantic drift: if increased embedding distance reflected off-topic wandering rather than meaningful novelty, we would expect an inverse relationship with coherence ratings. Yet half of participants feared coherence trade-offs---an expectation-reality gap suggesting intuitive mental models where creativity costs readability, even when evidence shows otherwise within the tested $\lambda$ range.

The marginal significance of perceived originality ($p = .062$) alongside significant usefulness ($p = .019$) suggests users recognize practical value in divergent outputs without reliably detecting originality in blind evaluation. Awareness of consensus may matter more than perceptual sensitivity---pointing toward design opportunities in making the positive originality-coherence relationship explicit to users.

\textbf{Design Implications.} Three qualitative findings carry design implications. First, only 25\% correctly identified the Yellow Zone as consensus clustering, while 38\% interpreted it as indicating optimal responses---revealing that users read visual density as quality rather than frequency. Future consensus visualizations should use explicit labeling (e.g., ``common'' vs.\ ``good'') and progressive disclosure to prevent this conflation.

Second, participants applied different evaluation criteria across task modes: accuracy in Technical writing, feasibility in Brainstorming, and vividness in Creative writing. This validates SRT's mode-specific configurations (different $\tau$, $\rho$, $n$ per mode) and suggests consensus-aware systems should surface mode-appropriate quality indicators alongside divergence metrics.

Third, several participants wanted consensus visibility for
manual editing rather than automated divergence---preferring
to see frequent combinations first then revise
themselves---suggesting SRT can also function as a diagnostic
tool with users retaining editorial control. System
preferences were elicited after the Yellow Zone explanation,
which may have primed participants toward consensus-aware
systems; these should be interpreted cautiously alongside
blind Likert ratings.

\textbf{Limitations.} Our sample (N=16) limits generalizability, and the visualization's mixed interpretability (25\% correct, 38\% conflating density with quality) motivates progressive disclosure in future work. Ablation studies are needed to isolate component contributions, and longitudinal studies to assess whether consensus awareness durably changes creative practice.


%
%

\appendix

\section{Experiment 2: Full Ablation Results}
\label{appendix:ablation}

This appendix reports the complete 8-condition ablation results across all
three modes, supplementing the main-effect delta summary in
Table~\ref{tab:ablation}. The 2$^3$ factorial design toggles Contrastive
Decoding~(CD), Phrase Penalties~(PP), and Fluency Controls~(FL) independently,
yielding eight conditions per mode. Conditions are labelled C0 (no mechanisms)
through C7 (all mechanisms enabled); the full condition legend is given in
Table~\ref{tab:ablation_legend}. All values are means over
$N=5$~samples~$\times$~10~prompts~$= 50$ observations per cell.

\begin{table}[!htb]
\centering
\small
\caption{Full ablation results --- Creative mode (mean, $N=50$).
Shaded row = Full SRT (C7).}
\label{tab:ablation_creative}
\setlength{\tabcolsep}{3.8pt}
\begin{tabular}{lccccccc}
\toprule
\textbf{Cond.} & \textbf{CD} & \textbf{PP} & \textbf{FL} &
\textbf{Orig.} & \textbf{Rel.} & \textbf{Cliché} & \textbf{PPL} \\
\midrule
C0 & \texttimes & \texttimes & \texttimes & 0.293 & 0.434 & 1.380 & 5.9 \\
C1 & \checkmark & \texttimes & \texttimes & 0.925 & 0.040 & 0.000 & 123{,}079 \\
C2 & \texttimes & \checkmark & \texttimes & 0.317 & 0.420 & 0.140 & 6.9 \\
C3 & \texttimes & \texttimes & \checkmark & 0.258 & 0.473 & 1.280 & 5.5 \\
C4 & \checkmark & \checkmark & \texttimes & 0.921 & 0.039 & 0.000 & 119{,}932 \\
C5 & \checkmark & \texttimes & \checkmark & 0.482 & 0.298 & 0.180 & 75.9 \\
C6 & \texttimes & \checkmark & \checkmark & 0.305 & 0.415 & 0.140 & 5.9 \\
\rowcolor{gray!15}
C7 & \checkmark & \checkmark & \checkmark & 0.484 & 0.299 & 0.040 & 71.3 \\
\bottomrule
\end{tabular}
\end{table}

\begin{table}[!htb]
\centering
\small
\caption{Full ablation results --- Technical mode (mean, $N=50$).
Shaded row = Full SRT (C7).}
\label{tab:ablation_technical}
\setlength{\tabcolsep}{3.8pt}
\begin{tabular}{lccccccc}
\toprule
\textbf{Cond.} & \textbf{CD} & \textbf{PP} & \textbf{FL} &
\textbf{Orig.} & \textbf{Rel.} & \textbf{Cliché} & \textbf{PPL} \\
\midrule
C0 & \texttimes & \texttimes & \texttimes & 0.054 & 0.722 & 0.000 & 2.0 \\
C1 & \checkmark & \texttimes & \texttimes & 0.812 & 0.171 & 0.000 & 88{,}739 \\
C2 & \texttimes & \checkmark & \texttimes & 0.060 & 0.713 & 0.000 & 2.0 \\
C3 & \texttimes & \texttimes & \checkmark & 0.055 & 0.716 & 0.000 & 2.1 \\
C4 & \checkmark & \checkmark & \texttimes & 0.829 & 0.170 & 0.000 & 90{,}267 \\
C5 & \checkmark & \texttimes & \checkmark & 0.149 & 0.662 & 0.000 & 9.2 \\
C6 & \texttimes & \checkmark & \checkmark & 0.056 & 0.715 & 0.000 & 2.1 \\
\rowcolor{gray!15}
C7 & \checkmark & \checkmark & \checkmark & 0.151 & 0.663 & 0.000 & 8.9 \\
\bottomrule
\end{tabular}
\end{table}

\begin{table}[!htb]
\centering
\small
\caption{Full ablation results --- Brainstorming mode (mean, $N=50$).
Shaded row = Full SRT (C7).}
\label{tab:ablation_brainstorm}
\setlength{\tabcolsep}{3.8pt}
\begin{tabular}{lccccccc}
\toprule
\textbf{Cond.} & \textbf{CD} & \textbf{PP} & \textbf{FL} &
\textbf{Orig.} & \textbf{Rel.} & \textbf{Cliché} & \textbf{PPL} \\
\midrule
C0 & \texttimes & \texttimes & \texttimes & 0.158 & 0.553 & 2.320 & 5.4 \\
C1 & \checkmark & \texttimes & \texttimes & 0.914 & 0.074 & 0.000 & 112{,}280 \\
C2 & \texttimes & \checkmark & \texttimes & 0.183 & 0.536 & 1.160 & 5.3 \\
C3 & \texttimes & \texttimes & \checkmark & 0.150 & 0.554 & 2.260 & 5.4 \\
C4 & \checkmark & \checkmark & \texttimes & 0.903 & 0.086 & 0.000 & 112{,}834 \\
C5 & \checkmark & \texttimes & \checkmark & 0.326 & 0.452 & 0.200 & 41.9 \\
C6 & \texttimes & \checkmark & \checkmark & 0.158 & 0.550 & 1.380 & 5.6 \\
\rowcolor{gray!15}
C7 & \checkmark & \checkmark & \checkmark & 0.355 & 0.454 & 0.220 & 40.0 \\
\bottomrule
\end{tabular}
\end{table}

\begin{table}[!htb]
\centering
\small
\caption{Ablation condition legend.}
\label{tab:ablation_legend}
\begin{tabular}{lccc}
\toprule
\textbf{Condition} & \textbf{Contrastive Decoding} &
\textbf{Phrase Penalties} & \textbf{Fluency Controls} \\
\midrule
C0 & \texttimes & \texttimes & \texttimes \\
C1 & \checkmark & \texttimes & \texttimes \\
C2 & \texttimes & \checkmark & \texttimes \\
C3 & \texttimes & \texttimes & \checkmark \\
C4 & \checkmark & \checkmark & \texttimes \\
C5 & \checkmark & \texttimes & \checkmark \\
C6 & \texttimes & \checkmark & \checkmark \\
C7 & \checkmark & \checkmark & \checkmark \\
\bottomrule
\end{tabular}
\end{table}

\begin{figure*}[!t]
\centering
\includegraphics[width=\textwidth]{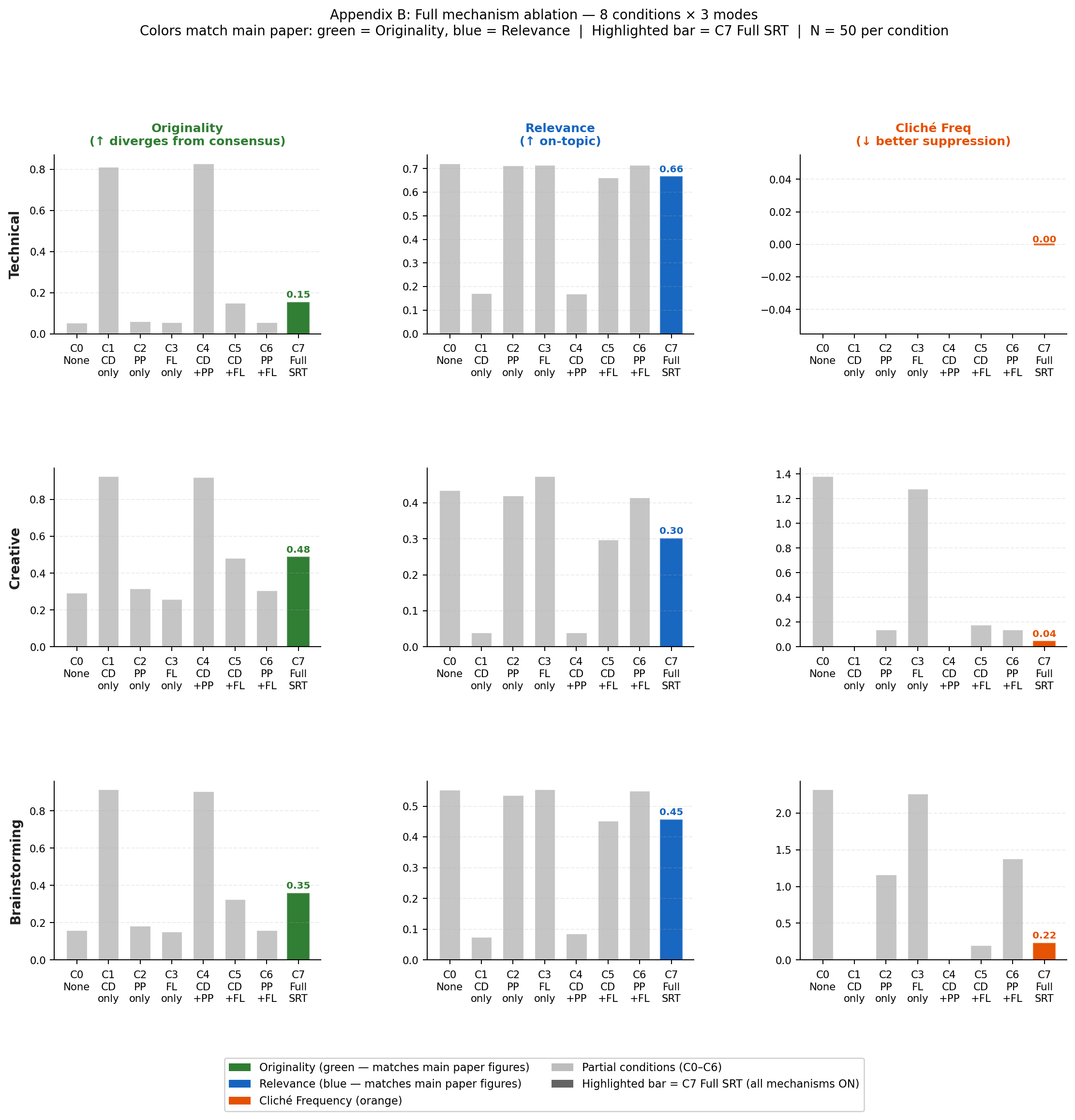}
\caption{Full ablation: 8 conditions $\times$ 3 modes for Originality,
Relevance, and Cliché Frequency. Highlighted bar (C7) = Full SRT.
C1 and C4 (CD active, FL inactive) achieve high originality but at the
cost of catastrophic coherence (PPL\,$>\,88{,}000$), confirming that
Fluency Controls are a necessary stabilizer rather than an optional component.}
\label{fig:ablation_full}
\end{figure*}

Figure~\ref{fig:ablation_full} visualises the full ablation across all
conditions and modes for Originality, Relevance, and Cliché Frequency.

\section{Experiment 3: Full $\lambda$-Sweep Results}
\label{appendix:lambda}

This appendix reports the complete per-mode $\lambda$-sweep results
supplementing the overall summary in Table~\ref{tab:lambda}. Values are
means~$\pm$ SD over $N=5$~samples~$\times$~10~prompts~$= 50$ observations
per cell. The Relevance and Coherence ($-$NLL) metrics characterize the
divergence--relevance--coherence trade-off across the full operating range.

\begin{table*}[!t]
\centering
\small
\caption{Per-mode $\lambda$-sweep results: mean $\pm$ SD ($N = 50$ per cell).
Orig.\ = Originality; Div.\ = Intra-Diversity; Cliché = Consensus Phrase
Frequency; Rel.\ = Relevance; Coh.\ = Coherence~($-$NLL); PPL = Perplexity.}
\label{tab:lambda_full}
\setlength{\tabcolsep}{4pt}
\begin{tabular}{llcccccc}
\toprule
\textbf{Mode} & $\lambda$ &
\textbf{Orig.} & \textbf{Div.} & \textbf{Cliché} &
\textbf{Rel.} & \textbf{Coh.} & \textbf{PPL} \\
\midrule
\multirow{5}{*}{Technical}
  & 0.0 & $0.096\pm0.084$ & $0.103\pm0.069$ & $0.000\pm0.000$ & $0.710\pm0.082$ & $-0.71\pm0.23$ & $2.1\pm0.5$ \\
  & 0.6 & $0.111\pm0.086$ & $0.127\pm0.066$ & $0.000\pm0.000$ & $0.696\pm0.086$ & $-1.13\pm0.46$ & $3.4\pm1.7$ \\
  & 1.2 & $0.179\pm0.099$ & $0.195\pm0.098$ & $0.000\pm0.000$ & $0.661\pm0.094$ & $-1.95\pm0.93$ & $10.2\pm8.6$ \\
  & 1.8 & $0.212\pm0.101$ & $0.201\pm0.095$ & $0.000\pm0.000$ & $0.653\pm0.095$ & $-2.35\pm1.10$ & $17.9\pm17.6$ \\
  & 2.4 & $0.215\pm0.081$ & $0.215\pm0.089$ & $0.000\pm0.000$ & $0.653\pm0.096$ & $-2.71\pm1.17$ & $26.5\pm26.4$ \\
\midrule
\multirow{5}{*}{Creative}
  & 0.0 & $0.293\pm0.088$ & $0.404\pm0.101$ & $0.200\pm0.404$ & $0.417\pm0.130$ & $-1.79\pm0.17$ & $6.1\pm1.0$ \\
  & 0.6 & $0.390\pm0.083$ & $0.489\pm0.090$ & $0.080\pm0.274$ & $0.332\pm0.120$ & $-3.01\pm0.22$ & $20.9\pm5.0$ \\
  & 1.2 & $0.466\pm0.099$ & $0.499\pm0.104$ & $0.060\pm0.240$ & $0.310\pm0.123$ & $-4.26\pm0.26$ & $73.2\pm19.3$ \\
  & 1.8 & $0.509\pm0.107$ & $0.465\pm0.093$ & $0.060\pm0.240$ & $0.274\pm0.127$ & $-4.73\pm0.29$ & $118.2\pm33.4$ \\
  & 2.4 & $0.500\pm0.103$ & $0.424\pm0.093$ & $0.020\pm0.141$ & $0.288\pm0.117$ & $-4.97\pm0.32$ & $150.9\pm46.3$ \\
\midrule
\multirow{5}{*}{Brainstorming}
  & 0.0 & $0.162\pm0.069$ & $0.242\pm0.074$ & $1.340\pm1.272$ & $0.551\pm0.097$ & $-1.60\pm0.33$ & $5.3\pm2.0$ \\
  & 0.6 & $0.257\pm0.083$ & $0.358\pm0.051$ & $0.260\pm0.487$ & $0.502\pm0.103$ & $-2.52\pm0.36$ & $13.2\pm4.9$ \\
  & 1.2 & $0.329\pm0.092$ & $0.399\pm0.051$ & $0.160\pm0.370$ & $0.472\pm0.091$ & $-3.63\pm0.55$ & $44.9\pm35.1$ \\
  & 1.8 & $0.374\pm0.078$ & $0.373\pm0.087$ & $0.080\pm0.274$ & $0.456\pm0.082$ & $-4.31\pm0.42$ & $80.5\pm32.3$ \\
  & 2.4 & $0.393\pm0.093$ & $0.415\pm0.076$ & $0.040\pm0.198$ & $0.434\pm0.117$ & $-4.49\pm0.46$ & $99.3\pm48.0$ \\
\bottomrule
\end{tabular}
\end{table*}

\begin{table*}[!htb]
\centering
\small
\caption{Marginal originality and relevance gains per $\lambda$ step
(averaged across modes, $N=150$). The near-zero gain at
$1.8 \to 2.4$ confirms the practical operating ceiling.}
\label{tab:marginal}
\setlength{\tabcolsep}{5pt}
\begin{tabular}{lrrr}
\toprule
$\lambda$ \textbf{step} & $\Delta$\textbf{Orig.} &
$\Delta$\textbf{Rel.} & $\Delta$\textbf{PPL} \\
\midrule
$0.0 \to 0.6$ & $+0.069$ & $-0.049$ & $+8.1$  \\
$0.6 \to 1.2$ & $+0.072$ & $-0.029$ & $+30.2$ \\
$1.2 \to 1.8$ & $+0.040$ & $-0.020$ & $+29.5$ \\
$1.8 \to 2.4$ & $+0.005$ & $-0.003$ & $+20.0$ \\
\bottomrule
\end{tabular}
\end{table*}

\begin{figure*}[!t]
\centering
\includegraphics[width=\textwidth]{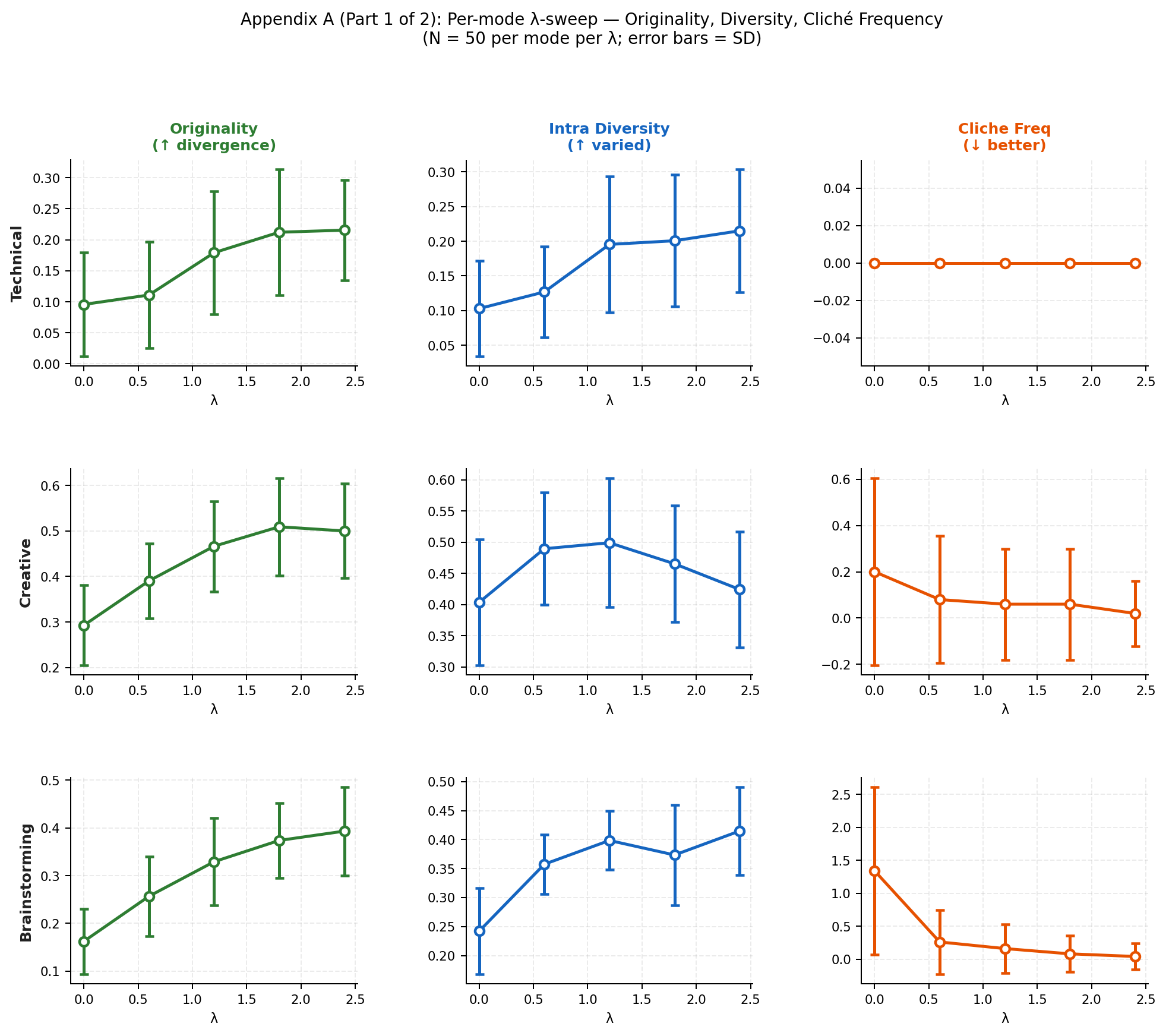}
\caption{Per-mode $\lambda$-sweep (Part~1 of~2): Originality
(green, $\uparrow$ divergence), Intra-Diversity (blue, $\uparrow$ varied),
and Cliché Frequency (orange, $\downarrow$ better suppression).
Rows correspond to modes; columns to metrics.
$N = 50$ per mode per $\lambda$; error bars = SD.}
\label{fig:per_mode_part1}
\end{figure*}

\begin{figure*}[!t]
\centering
\includegraphics[width=\textwidth]{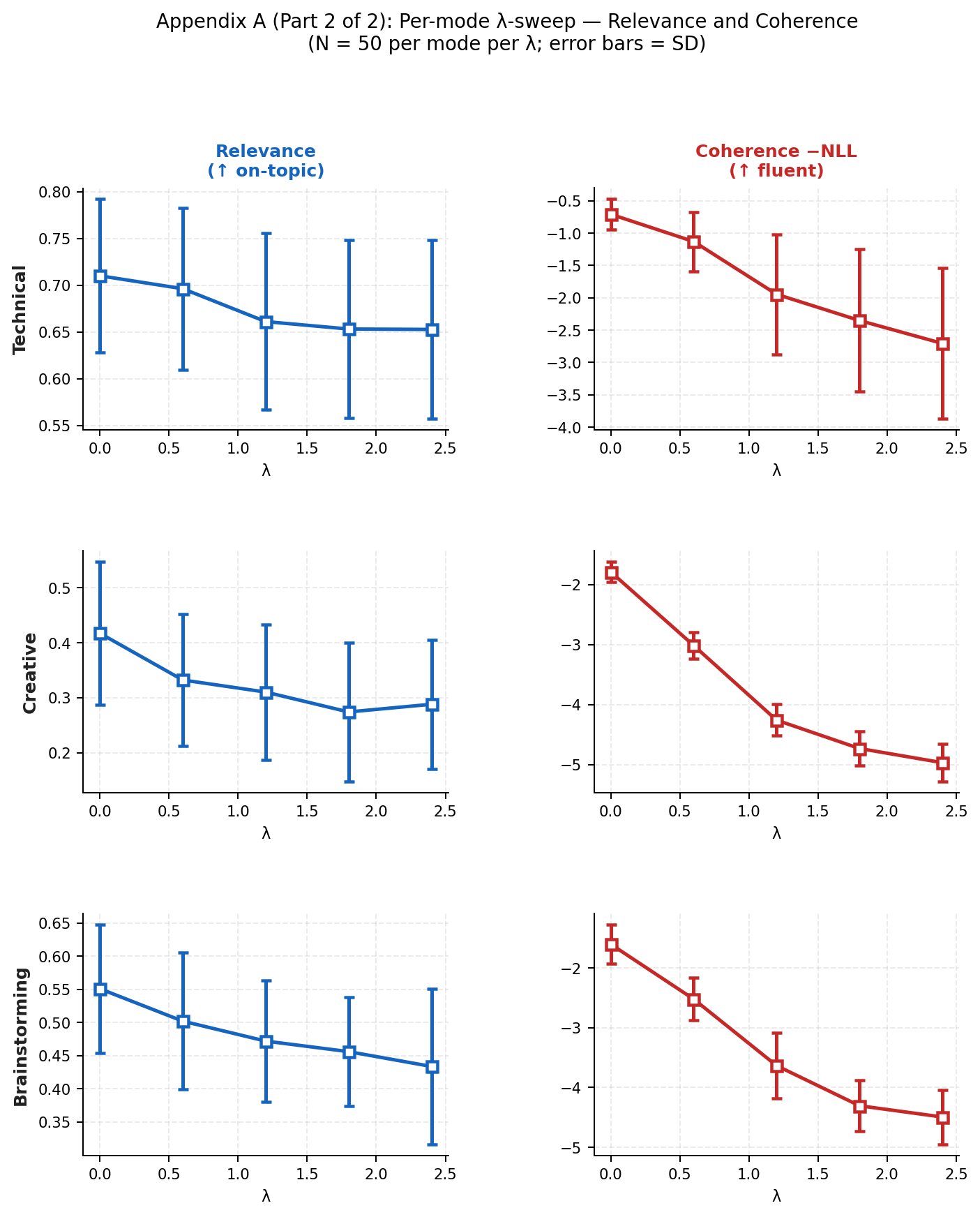}
\caption{Per-mode $\lambda$-sweep (Part~2 of~2): Relevance
(blue, $\uparrow$ on-topic) and Coherence~$-$NLL (red, $\uparrow$ fluent).
Technical mode (top row) retains relevance of $0.653$ even at
$\lambda = 2.4$, an $8.1\%$ drop from baseline, validating its lower
default configuration. Creative mode (middle row) shows the steepest
relevance decline ($30.9\%$), with the originality--relevance crossover
at $\lambda \approx 0.8$.}
\label{fig:per_mode_part2}
\end{figure*}

\begin{figure*}[!t]
\centering
\includegraphics[width=\textwidth]{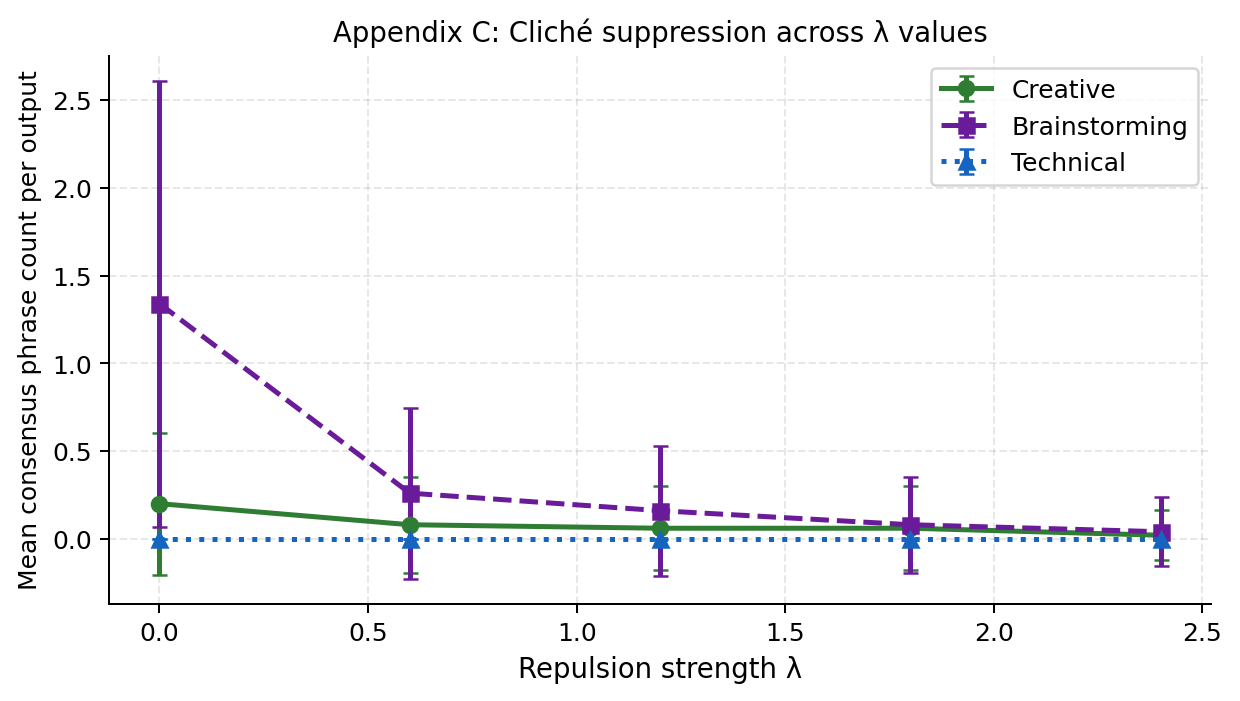}
\caption{Cliché suppression by mode. Brainstorming begins with the
highest baseline cliché rate ($1.34$ at $\lambda = 0$) due to marketing-frame
prevalence in LLM brainstorming outputs, and reaches near-zero by
$\lambda = 1.2$. Technical mode shows zero cliché frequency throughout
because the curated boilerplate phrases were not present in these
particular prompts' outputs---a finding that motivates extending the
Technical mode phrase list in future work.}
\label{fig:cliche_supp}
\end{figure*}

Figures~\ref{fig:per_mode_part1} and~\ref{fig:per_mode_part2} show all five
metrics per mode across the full $\lambda$ range.
Figure~\ref{fig:cliche_supp} shows cliché suppression separately by mode.

\section{Metric Definitions}
\label{appendix:metrics}

For reproducibility, we define all six metrics used across Experiments~1--3.

\begin{description}

\item[Originality.] Cosine distance between output embedding $\mathbf{e}_g$
and consensus centroid $\mathbf{c}$:
$D_{\text{mode}}(g) = 1 - \mathbf{e}_g^\top \mathbf{c}$.
Higher = more divergent from consensus. Embeddings are 384-dimensional
L2-normalised vectors from \texttt{all-MiniLM-L6-v2}.

\item[Intra-Diversity.] Mean pairwise cosine distance between all outputs
from the same system for the same prompt. Higher = more internally varied.

\item[Cliché Frequency.] Count of consensus-characterizing phrases
(extracted by YAKE~\cite{campos2020yake} for Creative, curated lists for
Technical and Brainstorming) appearing in each output. Lower = better
consensus suppression.

\item[Relevance.] Cosine similarity between the prompt embedding and the
output embedding: $\text{sim}(\mathbf{e}_{\text{prompt}},
\mathbf{e}_{\text{output}})$. Higher = more on-topic.

\item[Coherence ($-$NLL).] Negative mean token log-likelihood of the output
under the strong model:
$-\frac{1}{|y|}\sum_t \log p_{\theta}(y_t \mid y_{<t})$.
Higher (less negative) = more fluent. Provides an objective, scalable
coherence proxy independent of human raters.

\item[Perplexity.] $\exp(\text{NLL})$, i.e., $\exp(-\text{Coherence})$.
Lower = more fluent. A perplexity of 5 is characteristic of normal fluent
text; values above $1{,}000$ indicate near-incoherent output.

\end{description}
\noindent\textbf{Note on embedding-based metrics.} Originality and Relevance
are both computed from the same \texttt{all-MiniLM-L6-v2} embeddings.
These metrics capture semantic distance at the sentence level but cannot
distinguish diversity of \textit{expression} (different words, same idea)
from diversity of \textit{ideas} (genuinely different concepts). Future work
should complement these metrics with expert annotation or structured
idea-level evaluation.
\begin{table*}[p]
\centering
\small
\caption{Wilcoxon signed-rank tests: each $\lambda$ vs.\ $\lambda=0$
baseline ($N=150$ per $\lambda$, Bonferroni corrected).
All comparisons: *** $p < 0.001$.}
\label{tab:wilcoxon}
\setlength{\tabcolsep}{4pt}
\begin{tabular}{lcccc}
\toprule
$\lambda$ \textbf{vs 0} & \textbf{Orig.} & \textbf{Rel.} &
\textbf{Cliché} & \textbf{Coh.} \\
\midrule
0.6 vs 0.0 & *** & *** & *** & *** \\
1.2 vs 0.0 & *** & *** & *** & *** \\
1.8 vs 0.0 & *** & *** & *** & *** \\
2.4 vs 0.0 & *** & *** & *** & *** \\
\midrule
\textit{Friedman} $\chi^2$ & 338.67 & 161.54 & 114.69 & 525.60 \\
\textit{p} & $<.001$ & $<.001$ & $<.001$ & $<.001$ \\
\bottomrule
\end{tabular}
\end{table*}
\section{Statistical Significance: $\lambda$-Sweep}
\label{appendix:stats}

Table~\ref{tab:wilcoxon} reports Wilcoxon signed-rank test results comparing
each $\lambda > 0$ to the $\lambda = 0$ baseline, with Bonferroni correction
for four simultaneous comparisons ($\alpha_{\text{corrected}} = 0.0125$).
The Friedman test across all five $\lambda$ values additionally confirms
overall differences for each metric. All effects are $p < 0.001$, confirming
that even the smallest tested step ($\lambda = 0.6$) produces statistically
reliable divergence and that the observed trends are not attributable to
sampling variance.

\begin{figure*}[t]
\centering
\includegraphics[width=0.7\textwidth]{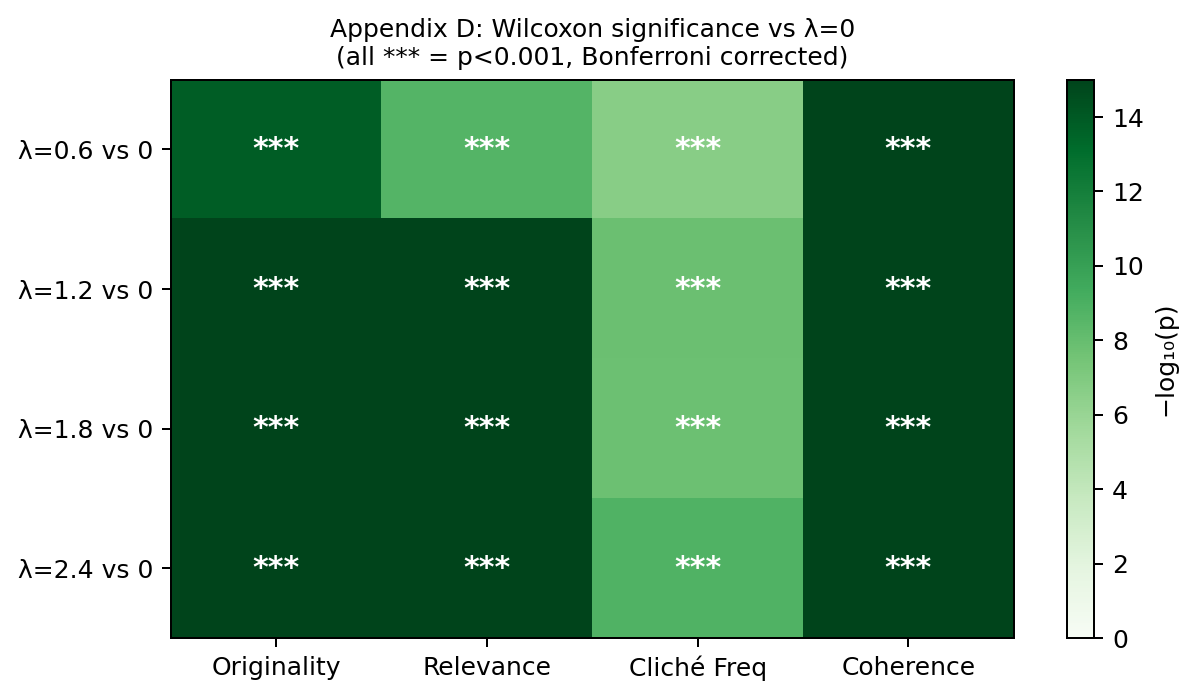}
\caption{Statistical significance heatmap. Color intensity encodes
$-\log_{10}(p)$; all cells show *** ($p < 0.001$), confirming that
every $\lambda$ step produces reliable, statistically significant effects
across all four metrics.}
\label{fig:significance}
\end{figure*}

\begin{algorithm*}[t]
\scriptsize
\setlength{\algorithmicindent}{1.1em}
\caption{Semantic Repulsion Technique (SRT)}
\label{alg:srt}
\begin{multicols}{2}
\begin{algorithmic}[1]
\Require Prompt $P$, task mode $M$, repulsion strength $\lambda$
\Require Strong model $S$, weak model $W$, embedding model $E$
\Ensure Repulsed response $Y$
\State $\mathcal{X} \gets \emptyset$
\Statex \textit{// Generate default responses to estimate consensus}
\For{$i = 1$ to $12$}
    \State $x_i \gets \textsc{Generate}(S, P, \text{temperature}=1.0, \text{top-}p=0.9)$
    \State $\mathcal{X} \gets \mathcal{X} \cup \{x_i\}$
\EndFor
\State $\mathcal{X} \gets \textsc{RemoveDuplicates}(\mathcal{X})$
\Statex \textit{// Compute semantic consensus centroid}
\ForAll{$x_i \in \mathcal{X}$}
    \State $e_i \gets \textsc{Normalize}(E(x_i))$
\EndFor
\State $c \gets \textsc{Normalize}\left(\sum_i e_i\right)$
\Statex \textit{// Extract consensus phrases to avoid}
\If{$M = \text{Creative}$}
    \State $\mathcal{N} \gets \textsc{ExtractRepeatedKeyphrases}(\mathcal{X}, P)$
\ElsIf{$M = \text{Technical}$}
    \State $\mathcal{N} \gets \textsc{FindBoilerplatePhrases}(\mathcal{X})$
\ElsIf{$M = \text{Brainstorm}$}
    \State $\mathcal{N} \gets \textsc{FindMarketingCliches}(\mathcal{X})$
    \If{$|\mathcal{N}| < 12$}
        \State $\mathcal{N} \gets \mathcal{N} \cup \textsc{ExtractRepeatedKeyphrases}(\mathcal{X}, P)$
    \EndIf
\EndIf
\Statex \textit{// Convert negative phrases into token penalties}
\State $\mathcal{T}_{\mathrm{neg}} \gets \emptyset$
\ForAll{$n \in \mathcal{N}$}
    \State $\mathcal{T}_{\mathrm{neg}} \gets \mathcal{T}_{\mathrm{neg}} \cup \textsc{Tokenize}(n)$
\EndFor
\Statex \textit{// Set mode-specific generation parameters}
\If{$M = \text{Creative}$}
    \State $\tau \gets 0.003$, $\beta \gets 4.0$, $\rho \gets 1.06$, $g \gets 4$, $L \gets 320$
\ElsIf{$M = \text{Technical}$}
    \State $\tau \gets 0.010$, $\beta \gets 4.0$, $\rho \gets 1.12$, $g \gets 6$, $L \gets 220$
\ElsIf{$M = \text{Brainstorm}$}
    \State $\tau \gets 0.002$, $\beta \gets 4.0$, $\rho \gets 1.06$, $g \gets 4$, $L \gets 260$
\EndIf
\Statex \textit{// Prepare decoding}
\State $V \gets \textsc{AlignVocabularies}(S, W)$
\State $Y \gets \textsc{Tokenize}(P)$
\Statex \textit{// Generate using semantic repulsion}
\For{$t = 1$ to $L$}
    \State $\ell_S \gets S.\textsc{NextTokenLogits}(Y)$
    \State $Y_W \gets \textsc{Tokenize}_{W}(\textsc{Detokenize}(Y))$
    \State $\ell_W^{\mathrm{raw}} \gets W.\textsc{NextTokenLogits}(Y_W)$
    \State $\ell_W \gets \textsc{AlignLogits}(\ell_W^{\mathrm{raw}}, V)$
    \Statex \textit{// Core semantic repulsion step}
    \State $\ell \gets \ell_S - \lambda \ell_W$
    \Statex \textit{// Penalize consensus tokens}
    \ForAll{$v \in \mathcal{T}_{\mathrm{neg}}$}
        \State $\ell[v] \gets \ell[v] - \beta$
    \EndFor
    \Statex \textit{// Preserve fluency}
    \State $p_S \gets \textsc{Softmax}(\ell_S)$
    \ForAll{token $v$}
        \If{$p_S[v] < \tau$}
            \State $\ell[v] \gets -\infty$
        \EndIf
    \EndFor
    \Statex \textit{// Avoid repetition}
    \State $\ell \gets \textsc{ApplyRepetitionPenalty}(\ell, Y, \rho)$
    \State $\ell \gets \textsc{BlockRepeatedNgrams}(\ell, Y, g)$
    \Statex \textit{// Sample next token}
    \State $\ell \gets \textsc{TopPFilter}(\ell, 0.9)$
    \State $y_t \gets \textsc{Sample}(\textsc{Softmax}(\ell))$
    \State $Y \gets Y \cup \{y_t\}$
    \If{$y_t = \textsc{EndOfText}$}
        \State \textbf{break}
    \EndIf
\EndFor
\State $Y \gets \textsc{Detokenize}(Y)$
\State $Y \gets \textsc{RemovePromptPrefix}(Y, P)$
\State \Return $Y$
\end{algorithmic}
\end{multicols}
\end{algorithm*}

\end{document}